\documentclass[sn-mathphys,Numbered]{sn-jnl}

\usepackage{graphicx}%
\usepackage{multirow}%
\usepackage{amsmath,amssymb,amsfonts}%
\usepackage{amsthm}%
\usepackage{mathrsfs}%
\usepackage[title]{appendix}%
\usepackage{xcolor}%
\usepackage{textcomp}%
\usepackage{manyfoot}%
\usepackage{booktabs}%
\usepackage{algorithm}%
\usepackage{algorithmicx}%
\usepackage{algpseudocode}%
\usepackage{listings}%
\usepackage{lscape}
\usepackage{ulem}

\theoremstyle{thmstyleone}%
\theoremstyle{thmstyletwo}%

\theoremstyle{thmstylethree}%

\begin{document}

\title[Two waves of massive stars running away from the young cluster R136]{Two waves of massive stars running away from the young cluster R136}


\author*[1]{\fnm{Mitchel} \sur{Stoop}}

\author[1,2]{\fnm{Alex} \sur{de Koter}}

\author[1]{\fnm{Lex} \sur{Kaper}}

\author[1]{\fnm{Sarah} \sur{Brands}}

\author[3]{\fnm{Simon} \sur{Portegies Zwart}}

\author[2]{\fnm{Hugues} \sur{Sana}}

\author[4]{\fnm{Fiorenzo} \sur{Stoppa}}

\author[5,6]{\fnm{Mark} \sur{Gieles}}

\author[7]{\fnm{Laurent} \sur{Mahy}}

\author[8]{\fnm{Tomer} \sur{Shenar}}

\author[1]{\fnm{Difeng} \sur{Guo}}

\author[4,9,2]{\fnm{Gijs} \sur{Nelemans}}

\author[2,10]{\fnm{Steven} \sur{Rieder}}


\affil*[1]{\orgdiv{Anton Pannekoek Institute for Astronomy}, \orgname{University of Amsterdam}, \orgaddress{\street{Science Park 904}, \city{Amsterdam}, \postcode{1098 XH}, \country{the Netherlands}}}

\affil[2]{\orgdiv{Institute of Astronomy}, \orgname{KU Leuven}, \orgaddress{\street{Celestijnenlaan 200D}, \city{3001 Leuven}, \country{Belgium}}}

\affil[3]{\orgdiv{Leiden Observatory}, \orgname{Leiden University}, \orgaddress{\street{P.O. Box 9513}, \city{Leiden}, \postcode{NL-2300 RA}, \country{the Netherlands}}}

\affil[4]{\orgdiv{Department of Astrophysics/IMAPP}, \orgname{Radboud University}, \orgaddress{\street{P.O. Box 9010}, \city{Nijmegen}, \postcode{NL-6500 GL}, \country{the Netherlands}}}

\affil[5]{\orgdiv{ICREA}, \orgaddress{\street{Pg. Llu\'{i}s Companys 23}, \city{Barcelona}, \postcode{E08010}, \country{Spain}}}

\affil[6]{\orgdiv{Institut de Ci\`{e}ncies del Cosmos (ICCUB)}, \orgname{Universitat de Barcelona (IEEC-UB)}, \orgaddress{\street{Mart\'{i} Franqu\`{e}s 1}, \city{Barcelona}, \postcode{E08028}, \country{Spain}}}

\affil[7]{\orgdiv{Royal Observatory of Belgium}, \orgaddress{\street{Avenue Circulaire 3}, \city{Brussels}, \postcode{B-1180}, \country{Belgium}}}

\affil[8]{\orgdiv{The School of Physics and Astronomy}, \orgname{Tel Aviv University}, \orgaddress{\street{P.O. Box 39040}, \city{Tel Aviv}, \postcode{6997801}, \country{Israel}}}

\affil[9]{\orgdiv{SRON}, \orgname{Netherlands Institute for Space Research}, \orgaddress{\street{Niels Bohrweg 4}, \city{Leiden}, \postcode{2333 CA}, \country{The Netherlands}}}

\affil[10]{\orgdiv{Geneva Observatory}, \orgname{University of Geneva}, \orgaddress{\street{Chemin Pegasi 51}, \city{Sauverny}, \postcode{1290}, \country{Switzerland}}}

\abstract{Massive stars are predominantly born in stellar associations or clusters \cite{deWit2005}. Their radiation fields, stellar winds, and supernovae strongly impact their local environment. In the first few million years of a cluster’s life, massive stars are dynamically ejected running away from the cluster at high speed \cite{Fujii2011}. However, the production rate of dynamically ejected runaways is poorly constrained. Here we report on a sample of 55 massive runaway stars ejected from the young cluster R136 in the Large Magellanic Cloud. Astrometric analysis with {\it Gaia} \cite{GaiaCollaboration2016,GaiaCollaboration2021,GaiaCollaboration2023} reveals two channels of dynamically ejected runaways. The first channel ejects massive stars in all directions and is consistent with dynamical interactions during and after the birth of R136. The second channel launches stars in a preferred direction and may be related to a cluster interaction. We find that 23-33\% of the most luminous stars initially born in R136 are runaways. Model predictions \cite{Fujii2011,Banerjee2012,Oh2015} have significantly underestimated the dynamical escape fraction of massive stars. Consequently, their role in shaping and heating the interstellar and galactic medium, along with their role in driving galactic outflows, is far more important than previously thought \cite{Andersson2020,Steinwandel2023}.
}




\maketitle
\section{Main text}
\label{sec:Results}
The Large Magellanic Cloud (LMC), a satellite galaxy of the Milky Way, hosts the Tarantula Nebula (30 Doradus), a region containing more than a thousand massive stars, formed in multiple bursts of star formation in the past several tens of Myr \citep{Schneider2018}. The most recent star formation episode in this region gave birth to the dense cluster core Radcliffe 136 (R136). With \textit{Gaia} Data Release 3 astrometric information \citep{GaiaCollaboration2016,GaiaCollaboration2021,GaiaCollaboration2023} we identify stars consistent with running away from R136 (see Methods). We require them to have a transverse velocity significantly larger than 27.6 km s$^{-1}$ and to be ejected up to 3 Myr ago, though R136 is probably younger. This yields 55 early-type runaways increasing the number of known runaways coming from the cluster core by an order of magnitude \citep{Evans2010,Lennon2018,Sana2022}. We determine their dynamic trace-back age (kinematic age), indicating how long ago they were ejected from R136, and cross-match them with the literature to obtain their stellar parameters (see Supplementary Information). 

\begin{figure*}
\centering
\includegraphics[width=0.99\linewidth]{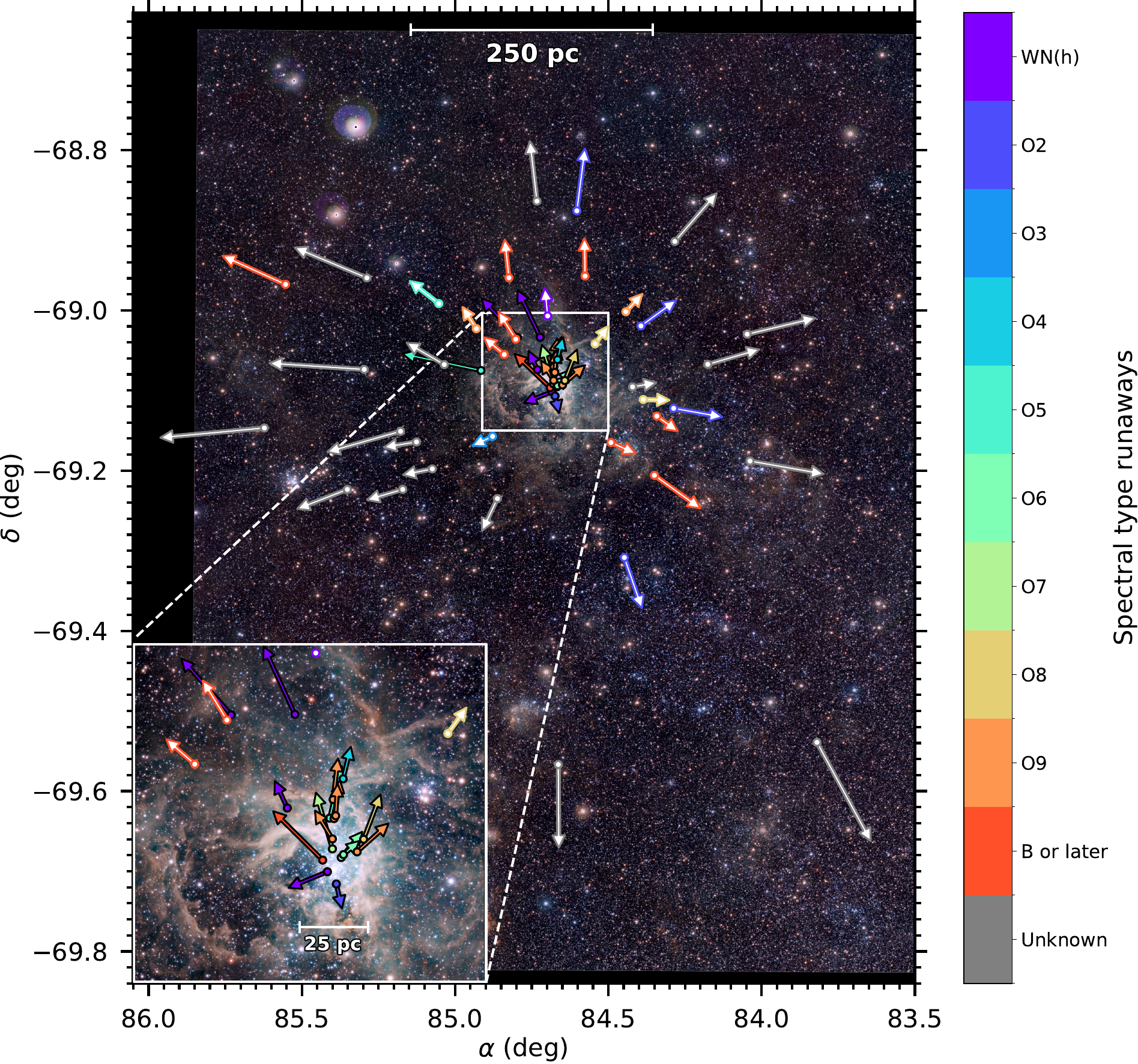}
\caption{\textbf{$\vert$ On-sky distribution of runaways coming from R136 in the last 3 Myr}. Arrows point into the direction of transverse motion; their length is proportional to the transverse velocity with respect to R136 (ranging from $\sim$ 28-195~km~s$^{-1}$). Black outlined and white filled markers and arrows indicate runaways with a kinematic age less and more than 1.0 Myr, respectively. The current position and transverse motion of the runaways are coloured according to spectral type. The background shows a near-infrared image of the Tarantula Nebula (VISTA, credit: ESO/M.-R. Cioni/VISTA Magellanic Cloud survey).}
\label{fig:runaways_onsky}
\end{figure*}

The spectral type of almost all classified runaways ranges from early-type B to early-type O, as well as WN(h)-type stars, with a corresponding mass in the range of $\sim$5 up to 140 M$_{\odot}$. The runaways move in different directions (Fig.~\ref{fig:runaways_onsky}), and have reached (projected) distances of $\sim$3 to 460 pc from R136. This implies that about half of them have left the 30~Dor region, and that their ionising radiation fields, supersonic stellar winds, and eventually powerful supernovae affect relatively tenuous areas in or outside the LMC. 

\renewcommand{\figurename}{Fig.}

\begin{figure}
\centering
\includegraphics[width=0.99\linewidth]{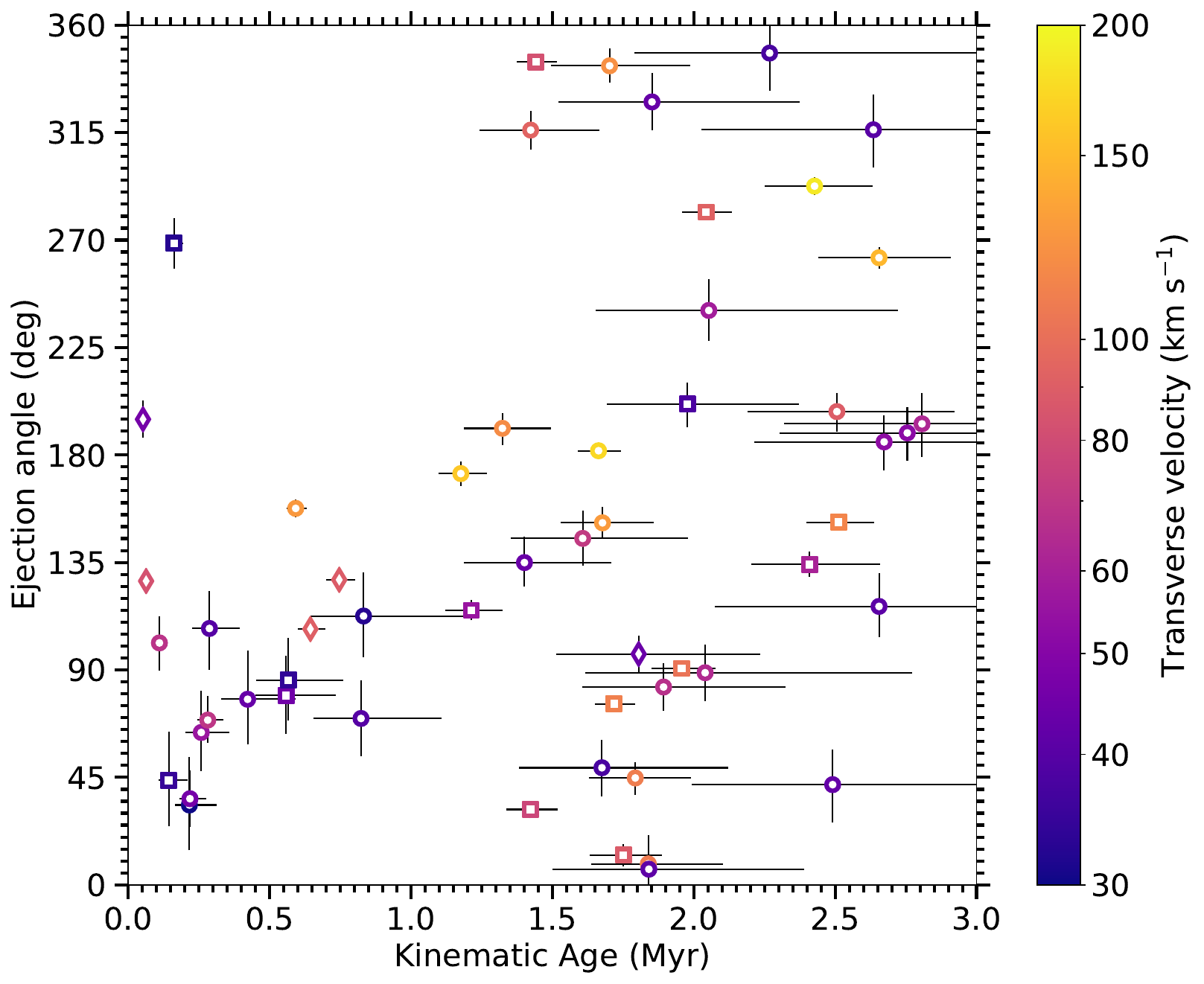}
\caption{\textbf{$\vert$ Ejection angle and kinematic age distribution of runaways coming from R136 in the last 3 Myr}. The markers are coloured according to their transverse velocity (logarithmic scale). Circles indicate runaways with 15 mag $<$ G $<$ 17 mag, squares with 13 mag $<$ G $<$ 15 mag, and diamonds the brightest runaways with 11 mag $<$ G $<$ 13 mag. The uncertainty on the data is expressed as a 1$\sigma$ confidence interval.}
\label{fig:r136_runaways_tkin_ejang}
\end{figure}

Remarkably, we find that the runaways are not consistent with being ejected isotropically (Fig.~\ref{fig:r136_runaways_tkin_ejang}), while we expect runaways to be ejected in a random direction via dynamical interactions between a binary and at least a single star \citep{Banerjee2012}. More specifically, the sub-sample of 18 runaways with kinematic age less than 1.0 Myr contains 16 runaways which were ejected to the north of R136. We can reject the hypothesis that the over-abundance of runaways ejected in a northern direction (PA = 20-150 deg) is due to an isotropic ejection mechanism to 7 $\cdot$ 10$^{-4}$ significance (see Supplementary Information). This indicates that these recent runaways are ejected in a preferred direction. This is not observed for the runaways with a kinematic age higher than 1.0~Myr. 

\renewcommand{\figurename}{Fig.}

\begin{figure*}
\centering
\includegraphics[width=0.99\linewidth]{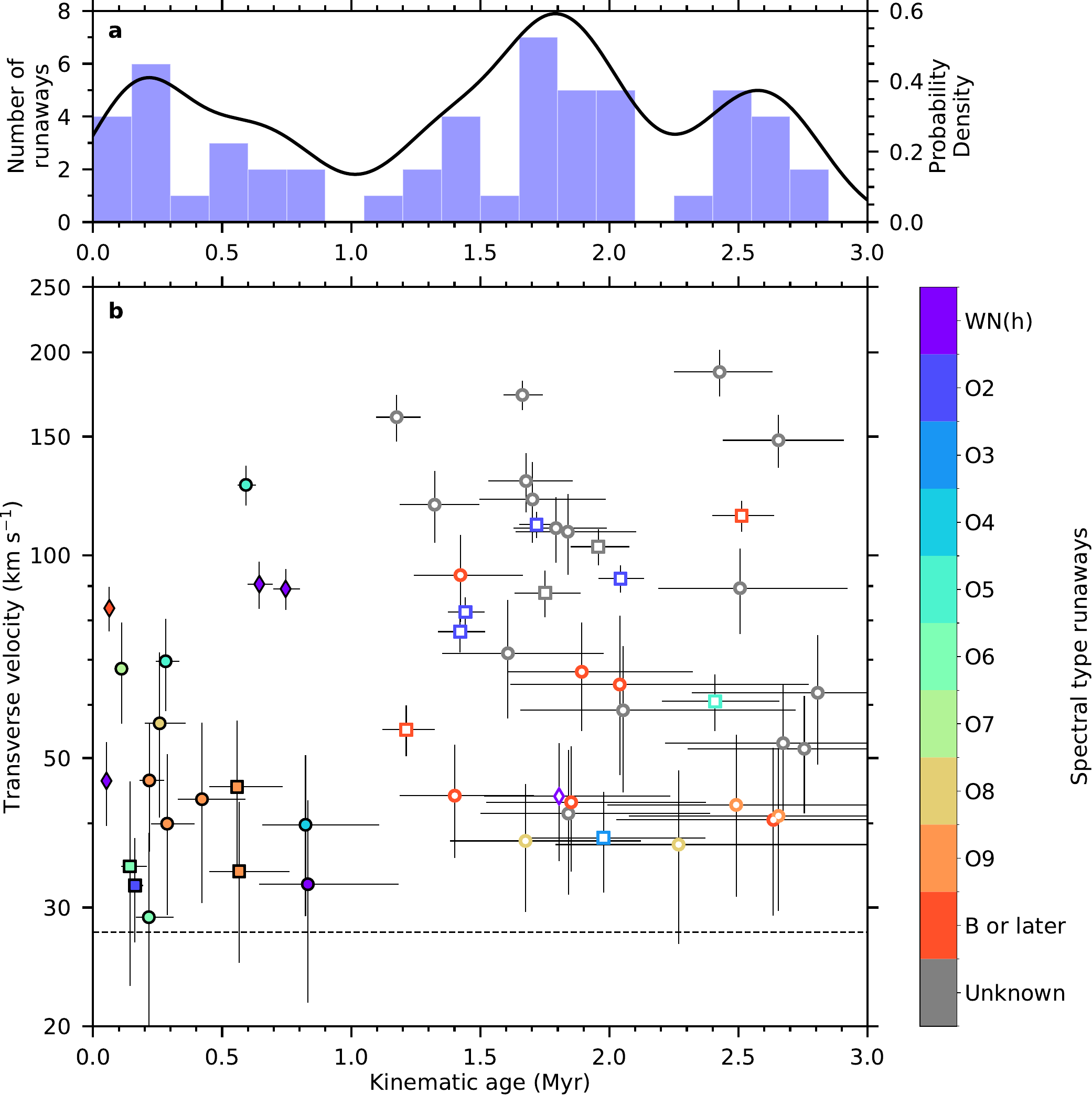}
\caption{\textbf{$\vert$ Kinematic age and transverse velocity distribution of runaways coming from R136 in the last 3 Myr}. \textbf{a}. Histogram of the kinematic ages in blue (left y-axis). The Gaussian kernel density estimator (bandwidth = 0.2 Myr) is shown with the black curve (right y-axis). \textbf{b}. The kinematic age and transverse velocity distribution of runaways coloured according to their spectral type. Circles indicate runaways with 15 mag $<$ G $<$ 17 mag, squares with 13 mag $<$ G $<$ 15 mag, and diamonds the brightest runaways with 11 mag $<$ G $<$ 13 mag. Black outlined and white filled markers indicate runaways ejected with a kinematic age less and greater than 1.0 Myr, respectively. The black dashed line shows the minimum transverse velocity required for a star to be classified as a runaway. The uncertainty on the data is given as a 1$\sigma$ confidence interval.}
\label{fig:r136_runaways_tkin_vt}
\end{figure*}

The kinematic age distribution in Fig.~\ref{fig:r136_runaways_tkin_vt} shows when the runaway stars were ejected. The intrinsic distribution of the kinematic ages is estimated using a Gaussian kernel density estimation (KDE) method. This reveals peaks around 0.2~Myr and 1.8~Myr; subsequently the distribution tailors off. The runaway ejection efficiency is lowest around $\sim$1.0~Myr. We conclude that the kinematic age distribution is not consistent with a constant ejection-rate to 2.5$\sigma$ significance. The peaks and dips in the observed KDE cannot be explained by stochastic sampling of the runaways. We interpret the peaks around 0.2~Myr and 1.8~Myr as separate ejection events, while few to no runaways were ejected about 1.0~Myr ago. The anisotropic nature of runaways with kinematic age less than 1.0 Myr relative to the isotropic nature of the runaways with larger kinematic age also points to two distinct ejection events.

The on-set of dynamical runaway ejection may be achieved during and shortly after the formation of a cluster \citep{Fujii2011,Oh2016}, as observed in several Galactic young massive clusters \citep{MaizApellaniz2022b_bermuda,Stoop2024}. The age of R136 is 1-2.5 Myr \citep{Brands2022}; therefore, the peak in the kinematic age distribution around $\sim$ 1.8 Myr ago can be identified as the on-set of the dynamical ejection of massive stars as part of the formation process of R136. Assuming that the runaway ejections in the last million years are not associated with this process, a Gaussian fit to the remainder of the sample yields a mean kinematic age for R136 of $\tau_{\rm{kin}}$ = 1.83$^{+0.14}_{-0.10}$ Myr. This provides an independent age estimate of R136, a benchmark star-bursting cluster, hosting the most massive stars known \citep{Crowther2010,Bestenlehner2020,Brands2022}.

\begin{figure*}
\centering
\includegraphics[width=0.99\linewidth]{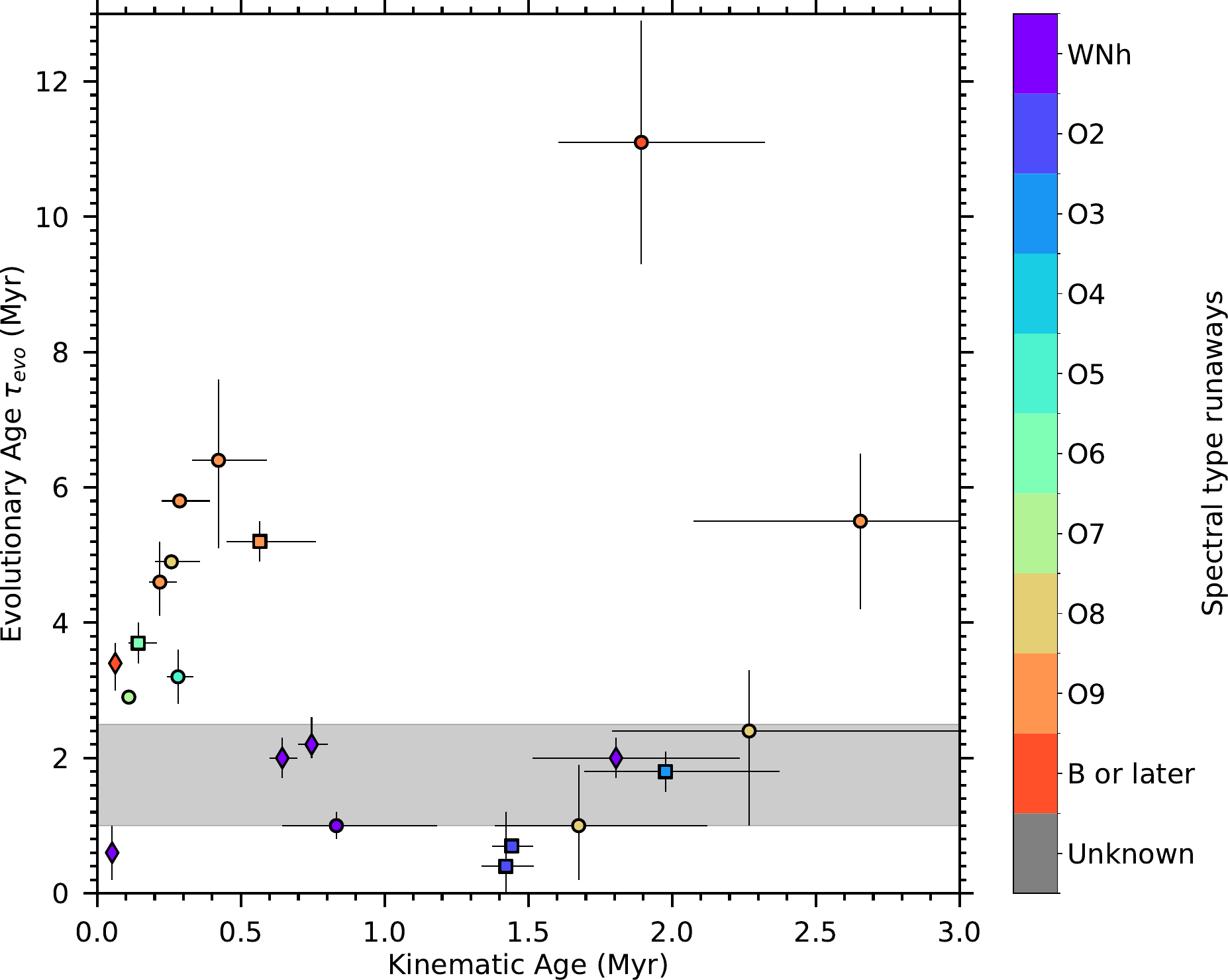}
\caption{\textbf{$\vert$ Distribution of the runaway evolutionary age and kinematic age}. The markers are coloured according to their spectral type. Circles indicate runaways with 15 mag $<$ G $<$ 17 mag, squares with 13 mag $<$ G $<$ 15 mag, and diamonds the brightest runaways with 11 mag $<$ G $<$ 13 mag. The grey band indicates the age of R136 \citep[1-2.5 Myr;][]{Brands2022}. The uncertainty on the data is given as a 1$\sigma$ confidence interval.}
\label{fig:runaways_tkin_tevo}
\end{figure*}

The evolutionary age has been estimated for 21 of the 55 runaways (Fig. \ref{fig:runaways_tkin_tevo}). Out of these 21, 11 have an evolutionary age larger than 2.5 Myr, which is unexpected given the evolutionary age of the stars in R136 \citep[1-2.5 Myr;][]{Brands2022}. The more evolved runaways have a median kinematic age of 0.28 Myr and thus have been ejected more recently. The evolutionary younger runaways have a median kinematic age of 1.43 Myr. We performed a K-means clustering analysis which supports the distinction between the runaways with an evolutionary age more and less than $\sim$ 2.0-2.5 Myr (see Supplementary Information).

\renewcommand{\figurename}{Fig.}

\renewcommand{\figurename}{Fig.}

The observation of these two different runaway populations suggests that two distinct dynamical ejection mechanisms are at play. The subset of runaways ejected longest ago can be explained by dynamical interactions in the centre of R136 during and shortly after the cluster formation. These runaways are consistent with being ejected isotropically and have an evolutionary age consistent with the age of R136. The runaways launched in a preferred direction are evolutionary older (3-7 Myr) and were ejected in a distinct episode only about 0.2 Myr ago, hence $\sim$ 1.5 Myr after the formation of R136. Here, we propose that these runaways were produced by an encounter of R136 with a known nearby cluster currently $\sim$ 5.4 pc to the north-east with an age of 2 to 5 Myr \citep{Sabbi2012}.


We estimate that 33\% of the stars born in R136 and more luminous than log($L$/L$_{\odot}$) $\gtrsim$ 6.0 (and more massive than $\sim$60~M$_{\odot}$) have been dynamically ejected from R136. The census of these very luminous stars both in the R136 core and in the surrounding field is complete, save for sources heavily obscured by interstellar dust \citep{Bestenlehner2014,Crowther2016}. Counting all non-runaway luminous field stars (see Supplementary Information) as originating from the core cluster as well would give a lower limit to the runaway fraction of 23\%. 

The runaway sample includes WN(h) and O-type stars, of which 30 are spectroscopically confirmed, and 4 are photometric candidates (G $<$ 16 mag). $N$-body simulations of young star clusters of similar mass as R136 predict significantly fewer ejections of such stars, typically by a factor of 2 to 6 \citep{Fujii2011,Banerjee2012,Oh2015}. A runaway fraction larger than 20\% is obtained only under special physical conditions: a high stellar density, strong primordial mass segregation, and a high primordial binary fraction. Possibly, a model which incorporates sub-cluster merging and the aforementioned starting conditions, but with a larger fraction of primordial wide binaries \citep{RamirezTannus2021}, could reproduce the observed massive star runaway fraction.


The more massive runaways initially born in R136 can (partially) explain the shallow slope ($\gamma_{\rm{30Dor}}$ = --1.90$^{+0.37}_{-0.26}$) of the stellar initial mass function, relative to the Salpeter slope of $\gamma = -2.35$ \citep{Salpeter1955}, determined in the surrounding 30 Doradus region, which does not include the dense core of R136 \citep{Schneider2018}. Excluding the five runaways with $M$ $>$ 60 M$_{\odot}$ also included in the field sample, steepens the mass-function slope to $\gamma_{\rm{30Dor}}$ $\sim$ --2.1, in marginal agreement with the Salpeter slope \citep{Salpeter1955}. Inside R136, the estimated slope of the mass function is $\gamma_{\rm{R136}}$ = --2.32 $\pm$ 0.16 \citep{Brands2022}. Correcting the cluster stars for the escaped runaways would make this shallower and we find $\gamma_{\rm{R136}}$ = --1.95 $\pm$ 0.08 for the stars with masses between 30-300 M$_{\odot}$. It is thus star formation in a dense cluster environment such as R136 that may produce a shallower slope of the high-mass end of the initial mass function.

The ionising radiation output of a young stellar population such as R136 is dominated by the most massive stars \citep{Doran2013}. The runaways found here contribute 22\% to the ionising budget of the most luminous stars (log$L$/L$_{\odot}$ $>$ 6.0) in 30 Doradus. Already 10\% of this budget is produced by runaways located outside the giant molecular cloud that defines 30 Doradus ($r_{\rm{proj}}$ $>$ 50 pc), providing a lower limit on the H-ionising photons able to escape the dusty star-forming environment. Luminous early-type stars in faint galaxies at redshifts $\sim 6-15$ are thought to constitute the dominant source of ionising radiation to explain the cosmological reionisation \citep{Barkana2001,Wise2009}. To reionise the universe early enough to be consistent with observations, the escape fraction of hydrogen ionising photons from these galaxies needs to be $f_{\rm{esc}}$ $\gtrsim$ 5-20\% \citep{Wise2009,Atek2023}. Simulations of photon escape from such early galaxies typically yield $f_{\rm esc} \sim 0.01-0.1$ \citep{Razoumov2007}, but do not take into account the effects of runaway stars \citep{Conroy2012}. If absorption by dust particles and atomic and molecular gas inside star-forming clouds is the dominant mechanism of Lyman-continuum photon breakdown, the so-far unaccounted for runaway stars may constitute the dominant source of $f_{\rm esc}$.


The ionising radiation and supernovae produced by massive early-type runaways cause efficient heating of the interstellar medium. Hydro-dynamical simulations of Milky Way-like spiral galaxies show that the inclusion of a runaway fraction of 14\% increases the heating of the inter-arm medium by an order of magnitude \citep{Andersson2020}. Runaway stars and their life ending core-collapse supernovae can also more efficiently (temporarily) expel gas and metals from their host galaxy, increasing the mass and metal loading factor of either a Milky Way-like or dwarf galaxy by a factor 2 to 10 \citep{Andersson2020,Steinwandel2023}, thus significantly affecting the dynamical and chemical evolution of a galaxy.


\clearpage
\section*{Methods}
\label{sec:method}
\subsection*{\textit{Gaia} data, filters, and corrections}
\label{sec:methods_gaia_filters}
We select stars in the field in and around 30 Doradus from the \textit{Gaia} Data Release 3 \citep[DR3;][]{GaiaCollaboration2016,GaiaCollaboration2021,GaiaCollaboration2023}. Taking sources in \textit{Gaia} DR3 within 1 deg from R136 with G $<$ 18 mag results in 83,295 sources. This radius is equivalent to $\sim$ 870 pc at the distance of the LMC \citep[49.59 kpc;][]{Pietrzynski2019} and sufficiently large to find runaways with a transverse velocity up to $\sim$ 290 km s$^{-1}$ ejected $\sim$ 3 Myr ago. The magnitude cut-off is to decrease computation time and the additional filters introduced below would remove sources with G $\gtrsim$ 17.5 mag regardless. A set of corrections and filters was first applied to the astrometry. The parallax zero-point offset, estimated from quasars, was used to correct for this bias \citep{Lindegren2021}. We have applied several filters to prevent spurious astrometric solutions from contaminating the runaway sample. First, the astrometry should have a 5 or 6-parameter astrometric solution. This excludes sources for which no proper motion and parallax could be found. Second, the renormalised unit weight error (\texttt{ruwe}) should be less than 1.4. A larger \texttt{ruwe} may indicate that the astrometric solution is spurious \citep{Lindegren2018}. This may bias our runaway sample against crowded regions and multiple systems. Third, the visibility periods adopted in the astrometric solution should be 10 or more (\texttt{visibility\_periods\_used}), designating the number of grouped observations used in the astrometric solution. If this is less than 10, it may indicate astrometric or photometric biases \citep{GaiaCollaboration2021}. Fourth, we remove sources for which the image parameter determination goodness of fit amplitude (\texttt{ipd\_gof\_harmonic\_amplitude}) is larger than 0.15. If this is the case, it may indicate that the astrometric solution is contaminated by crowding. Similarly we remove sources for which more than one peak was identified (\texttt{ipd\_frac\_multi\_peak}) in more than 10\% of the windows used by \textit{Gaia}, above which crowding may cause issues in the astrometric solution \citep{GaiaCollaboration2021}. Last, sources were excluded for which more than one source identifier was used in the data processing (\texttt{duplicated\_source}), which may indicate issues in the astrometric solution. From the initially 83,295 selected sources we are left with 71,391 sources.

To select sources consistent with being located in the LMC, we remove sources with a parallax $\varpi_{\rm{i}}$ larger than 0.15 mas and a parallax uncertainty $\sigma_{\varpi_{i}}$ larger than 0.05 mas. At the distance of 49.59 kpc for the LMC \citep[$\varpi_{\rm{LMC}}$ = 0.020 mas;][]{Pietrzynski2019}, the parallax should be less than $\varpi_{\rm{LMC}}$ + 3$\sigma_{\varpi,i}$ for each individual source. This excludes over two-third of the sources, leaving 21,382 sources. As a result of the parallax constraints, the faintest source has G $\sim$ 17.4 mag, implying that the latest / least massive runaway we can detect at the distance of the LMC is a B1-3 V star, with a mass of 5-10 M$_{\odot}$ \citep{Pecaut2013}.

\subsection*{Searching for runaways}
\label{sec:methods_runaway_search}
We search for massive runaways originating from the centre of R136. To do this, the remaining 21,382 stars in the field in and around 30 Doradus are investigated to identify whether they coincide with the position of R136, taking into account the proper motion of R136 itself. A classical runaway should have a velocity larger than 30 km s$^{-1}$. The 3D escape velocity of R136 at a distance of 1 pc from the centre, for a cluster mass of 5 $\times$ 10$^{4}$ M$_{\odot}$, is $\sim$ 21 km s$^{-1}$ \citep{Crowther2016}. The radial-velocity dispersion of single massive stars in and around R136 is $\sim$ 3.9 km s$^{-1}$, which has been corrected for the radial motion of undetected binaries \citep{HenaultBrunet2012}. For the definition of a runaway, we adopt a transverse velocity difference larger than 5$\sigma_{\rm{2D}}$ $\simeq$ 27.6 km s$^{-1}$, where $\sigma_{\rm{2D}}$ = $\sqrt{2}$$\sigma_{\rm{1D}}$. This is similar to the runaway threshold velocity in \cite{Sana2022}, who require a radial velocity deviating from the mean larger than three times the radial-velocity dispersion of all apparently single VFTS O stars in the 30 Doradus region ($\sim$ 25.8 km s$^{-1}$). Since this is a 1D velocity, a difference of factor $\sim$ $\sqrt{2}$ exists. Still undetected multiple systems could exist among the apparent single stars, which could increase the radial velocity dispersion. Runaways should not only have a transverse velocity significantly differing from that of R136, but their $v_{\rm{T}}$ should also be accurately known. The fractional uncertainty on the transverse velocity $v_{\rm{T}} / \sigma_{v_{\rm{T}}}$ should be more than 3, so that $v_{\rm{T}}$ is constrained.

We search for runaways by tracing back the stars using their proper motion. Since the R136 cluster also has a proper motion, we subtract this from each runaway candidate. Runaways are traced back following the equation
\begin{align*}
l_{\rm{sep,i}}(t) &= (l_{\rm{i}} + \frac{t \cdot \mu_{l^{*}_{\rm{i}},\rm{}}}{3.6 \times 10^{6} \cdot \rm{cos} (b_{\rm{i}})}) \\
&\quad - (l_{\rm{R136}} + \frac{t \cdot \mu_{l^{*},\rm{R136}}}{3.6 \times 10^{6} \cdot \rm{cos}(b_{\rm{R136}})})\ \rm{deg}, \\
b_{\rm{sep,i}}(t) &= (b_{\rm{i}} + \frac{t \cdot \mu_{b_{\rm{i}}}}{3.6 \times 10^{6}}) \\
&\quad - (b_{\rm{R136}} + \frac{t \cdot \mu_{b,\rm{R136}}}{3.6 \times 10^{6}})\ \rm{deg},
\end{align*}
where $t$ is the time in years. $l_{\rm{sep}}$ and $b_{\rm{sep}}$ give the separation of each runaway candidate with respect to the centre of R136. Galactic coordinates have been adopted to minimise the contribution of the cos($b$) factor. The runaway candidates have uncertainties on their proper motion, which is on average 0.045 mas yr$^{-1}$. This results in an average 1$\sigma$ uncertainty of $\sim$ 0.02 deg after 1.5 Myr. The radius of R136 is approximately two orders of magnitude smaller (3 $\times$ 10$^{-4}$ deg adopting a physical radius of 0.3 pc) than the semi-major axis of the 1$\sigma$ uncertainty ellipse, indicating that we can only trace back runaways to the approximate surroundings of R136. The position of the runaway candidates should be consistent within 2$\sigma$ at any time $t$ with that of the centre of R136 following
\begin{align}
\label{eq:methods_runaway_separation}
\sqrt{l^{2}_{\rm{sep,i}}(t) + b^{2}_{\rm{sep,i}}(t)} < \frac{2t}{3.6\times10^{6}}\ \rm{max}(\sigma_{\mu_{l^{*}_{\rm{i}}}},\sigma_{\mu_{b_{\rm{i}}}})\ \rm{deg},
\end{align}
where we take the maximum between $\sigma_{\mu_{l^{*}}}$ and $\sigma_{\mu_{b}}$. This indicates that the 2$\sigma$ uncertainty ellipse should overlap with the centre of R136 and therefore the position of the runaway is consistent within 2$\sigma$ to originate from the centre of R136. We trace back the candidate runaways up to 3 Myr ago given the maximum age of R136 \citep[1-2.5 Myr;][]{Brands2022}.

\subsection*{Proper motion of R136}
\label{sec:methods_r136_pm}
To identify runaways, we require prior information on the proper motion of R136. Since R136 is a highly crowded region, \textit{Gaia} resolved no stars with reliable astrometry within several parsec of the cluster centre. To illustrate this, we show in Extended~Data~Fig.~\ref{fig:r136_good_astrometry} the proper motion of stars relative to R136 with reliable astrometry as defined above, but which may not necessarily be runaways. Clearly no stars in R136 itself (black circle indicates a radius of 2.0 pc) have reliable astrometry. Information on the proper motion of R136 therefore must come from the surroundings, which may possibly contain runaways and is known to contain a north-east cluster \citep{Sabbi2012}. A simple approach is to take the mean or median of the proper motion of these stars in the surrounding, assuming that all proper motions are randomly oriented with respect to R136 and will average out. In this way, \cite{Sana2022} estimate $\mu_{\alpha^{*},\rm{R136}}$ = 1.700 mas yr$^{-1}$ $\mu_{\delta,\rm{R136}}$ $\simeq$ 0.684 mas yr$^{-1}$ and \cite{Lennon2018} estimate $\mu_{\alpha^{*},\rm{R136}}$ = 1.739 mas yr$^{-1}$ and $\mu_{\delta,\rm{R136}}$ = 0.701 mas yr$^{-1}$. 

If the proper motion of the stars surrounding R136 is not random (due to the orbit the north-east sub-cluster), this assumption could break down. Instead, we could use the runaways themselves to determine the proper motion of R136. A runaway ejected from the centre of R136 through dynamical interactions conveys information on where R136 was in the past and therefore what the proper motion of R136 is. We keep both $\mu_{l^{*},\rm{R136}}$ and $\mu_{b,\rm{R136}}$ as free parameters in our runaway search and adopt an iterative approach. This method has also been applied to the young massive cluster NGC\,6618 in M\,17 to show that the O-type runaways all have been ejected from within 0.1-0.2 pc from the cluster centre \citep{Stoop2024}.

For an initial R136 proper motion we find all runaways satisfying the conditions listed earlier. With these runaways we can redetermine the R136 proper motion which minimises the total $\sqrt{l^{2}_{\rm{sep, min}}(t) + b^{2}_{\rm{sep, min}}(t)}$, normalised to the total number of runaways found. As the uncertainties on the proper motion of the runaways differ, we normalise the minimum separation of each runaway with respect to R136 by its uncertainty on the position (the right-hand side of Equation~\ref{eq:methods_runaway_separation}). Field star interlopers could contaminate the runaway sample; we therefore only use candidate runaways with Bp - Rp $<$ 1.0 mag, G $<$ 15 mag, and $t_{\rm{kin}}$ $<$ 2.25 Myr to determine the R136 proper motion. These stars are consistent with bright early-type stars (see Section~\ref{sec:methods_fieldstars}), of which most have been classified as O or WN(h)-type stars. The candidate runaways should have $v_{\rm{T}}$ $>$ 25 km s$^{-1}$, $v_{\rm{T}}/\sigma_{v_{\rm{T}}}$ $>$ 2.5 and $t_{\rm{kin}}$ $<$ 3.0 Myr, which we calculate from the observed proper motion and uncertainties ignoring correlations. Having redetermined the R136 proper motion, we again search for the runaways. These two steps are iterated until the number of runaways and the R136 proper motion remain unchanged. A total of six iterations were required, after which 69 candidate runaways were found (with field stars removed).

After obtaining the candidate runaways, we determine $v_{\rm{T}}$ and $t_{\rm{kin}}$ with Markov Chain Monte Carlo simulations (MCMC, see below) and keep the runaways with $v_{\rm{T}}$ $>$ 27.6 km s$^{-1}$, $v_{\rm{T}}/\sigma_{v_{\rm{T}}}$ $>$ 3, and $t_{\rm{kin}}$ $<$ 3.0 Myr. This gives a final runaway sample of 55 runaways, which are assumed to be ejected from the centre of R136. We list the astrometric, kinematic, and physical parameters of R136 in Extended Data Table~\ref{tab:R136params}.

\subsection*{Field stars}
\label{sec:methods_fieldstars}
The candidate runaways will include field star interlopers. We are expecting bright early-type stars as runaways considering the faintest candidate runaway has G $\sim$ 17.4 mag, however, late-type red giants contaminate the runaway sample. The typical ages of red giants exceed the age of R136 by 2 to 3 orders of magnitude. They are considered interlopers. We can distinguish the early-type stars from red giants in colour-colour diagrams such as the \textit{2MASS} (J -- H) - (H -- K$_{\rm{s}}$) diagram in Extended~Data~Fig.~\ref{fig:runaways_jh_hk} \citep{Skrutskie2006}. Early-type stars should have (J -- H) $\lesssim$ 0 mag and (H -- K$_{\rm{s}}$) $\lesssim$ 0 mag, while red giants have (J -- H) $\gtrsim$ 0.5 mag. Interstellar extinction will cause both colours to be reddened and we show typical reddening lines for an early-type O9 V star and a late-type M0 V star with the black dashed lines (for a total-to-selective extinction R$_{\rm{V}}$ = 3.1). Even considering interstellar extinction, red giants are located in the upper-left part of this diagram, while early-type stars should be on the O-type reddening line. We have excluded stars consistent with the late-type part of this diagram indicated with the grey region. The reddened runaway VFTS 682, with spectral type WN5h, is shown with the blue star \citep[with extinction A$_{\rm{V}}$ = 4.45 $\pm$ 0.45;][]{Bestenlehner2011}, which overlaps with the O9 V reddening line.

The \textit{2MASS} JHK$_{\rm{s}}$ photometry is not available for nearly half of the stars with G $>$ 15 mag. We show the \textit{Gaia} colour magnitude diagram in Extended~Data~Fig.~\ref{fig:runaways_g_bprp} and the previously determined early-type stars with the blue circles and the late-type stars with the red squares. Candidate runaways with unavailable or poor \textit{2MASS} photometry are shown with the open symbols. Almost all stars with G$_{\rm{Bp}}$ -- G$_{\rm{Rp}}$ $\gtrsim$ 1.0 mag are determined to be late-type stars from the near-infrared colour-colour diagram, except for two. One of these is VFTS 682, while we classify the second star (purple open square) as a red giant and the 33 stars with G$_{\rm{Bp}}$ -- G$_{\rm{Rp}}$ $<$ 1.0 mag as early-type stars (golden open circles). A reasonable assumption is that early-type stars should have G$_{\rm{Bp}}$ -- G$_{\rm{Rp}}$ $<$ 1.0 and only VFTS 682 is found to be the exception to this. VFTS 682 is a well-studied WN(h) star \citep{Bestenlehner2011}, already identified as a potential runaway from R136 \citep{Renzo2019vfts682}. We have excluded the red supergiant MH 18 located in the top-right in Extended~Data~Fig.~\ref{fig:runaways_g_bprp} ($v_{\rm{T}}$ $\sim$ 60 km s$^{-1}$).

\subsection*{Transverse velocity and kinematic age}
\label{sec:methods_uncertainties}
We determine the transverse velocity and the kinematic age of each runaway with MCMC simulations. For each runaway, the three observables are ($\varpi_{i}$, $\mu_{l^{*},i}$, $\mu_{b,i}$) with a corresponding 3 by 3 covariance matrix. The three random variables are the distance $d_{i}$, transverse velocity $v_{\rm{T}}$$_{,i}$ and ejection angle $\phi_{i}$. The distance prior $\theta(d_{i})$ is a Gaussian distance distribution with a mean of 49.59 kpc and a standard deviation of 0.011 kpc \citep{Pietrzynski2019}. We adopt a uniform prior on the transverse velocity $\theta(v_{\rm{T}}$$_{,i})$ between 0-250 km s$^{-1}$ and ejection angle $\theta(\phi_{i})$ between 0-2$\pi$ rad. From this we calculate the three variables 
\begin{align*}
    \varpi_{i} &= 1 / d_{i}\ \rm{mas}, \\
    \mu_{l^{*},i} &= \frac{v_{\rm{T},\textit{i}} \cdot \rm{sin}(\phi_{\textit{i}})}{4.74047 \cdot d_{i}}\ \rm{mas}\ \rm{yr}^{-1},\\
    \mu_{b,i} &= \frac{v_{\rm{T},\textit{i}} \cdot \rm{cos}(\phi_{\textit{i}})}{4.74047 \cdot d_{i}}\ \rm{mas}\ \rm{yr}^{-1}.\\
\end{align*}
For each runaway, we draw four chains each containing 100,000 iterations to determine the posterior samples. With the inference data, the kinematic age can be determined as the time which minimises the separation between the runaway and R136. Assuming that the runaways are ejected from R136, we use the iterations for the kinematic age for which the minimum separation between the runaway and R136 is less than 0.005 deg ($\sim$ 5 pc at the distance of the LMC). The $v_{\rm{T}}$ and $t_{\rm{kin}}$ of a runaway are given by the 50$^{\rm{th}}$ percentiles, with a positive and negative 1$\sigma$ uncertainty determined by the 13.6$^{\rm{th}}$ and 86.4$^{\rm{th}}$ percentile, respectively.

\subsection*{Luminous stars in 30 Doradus}
We have collected all luminous stars in 30 Doradus with log($L$/L$_{\odot}$) $>$ 6.0 (Extended~Data~Table~\ref{tab:luminous_stars}). This includes both stars found inside the core of R136, the surrounding halo around R136, and in the 30 Doradus Nebula \citep{Bestenlehner2014,Schneider2018,Brands2022}. We have included MCPS 084.44781-69.30846 and SK-68 137, both of which are O2-3 stars consistent with log($L$/L$_{\odot}$) $>$ 6.0, outside the 30 Doradus region and for which no stellar parameters have been determined. For these two stars and VFTS 512, we take their stellar parameters as the average of the luminous stars in R136 with a similar spectral type.

We determine the runaway fraction (RF) of the stars with log($L$/L$_{\odot}$) as
\begin{align*}
    \rm{RF} = \frac{\rm{R}}{\rm{CL} + \rm{R}}\ \cdot 100\% = 33\%,
\end{align*}
where R ($n$ = 12) is the total number of runaway stars, and CL ($n$ = 25) is the total number of cluster stars. We define stars part of R136 that have a projected distance with respect to R136 $r_{\rm{proj}}$ $<$ 10 pc, except for Melnick 34 and VFTS 512 which we determine to be runaways. Field stars (FLD) are found in 30 Doradus which are neither runaways nor cluster stars ($n$ = 15). If we assume that these field stars may still originate from R136, but did not reach the escape velocity, they may be on wider elliptical orbits. Several field stars have \texttt{ruwe} $>$ 1.4, making it impossible to draw conclusions on their origin. We can also calculate the runaway fraction as
\begin{align*}
    \rm{RF} = \frac{\rm{R}}{\rm{CL} + \rm{FLD} + \rm{R}}\ \cdot 100\% = 23\%,
\end{align*}
providing a lower limit to the runaway fraction \citep[similar to what was found for Galactic O stars;][]{CarreteroCastrillo2023}. Extended~Data~Fig.~\ref{fig:dor30_log6_luminous} shows the location of the runaway, cluster and field stars in and outside the 30 Doradus Nebula. If we have missed runaways nearby R136 either because (systemic) radial velocities have not been determined or because the \textit{Gaia} astrometry is not accurate, the minimum runaway fraction may be higher. A radial-velocity search for O-type runaways in the 30 Doradus region yields at least one candidate with an evolutionary age less than 3.0 Myr and a mass $\gtrsim$ 60 M$_{\odot}$ \citep{Sana2022}.

\subsection*{Data and Code availability}
The datasets generated during and/or analysed during the current study are available in the following Zenodo repository once published, \url{https://doi.org/10.5281/zenodo.10058762}. The software and code developed in this manuscript to produce the findings in this study are available in the same repository.

\bibliography{MAIN.bib}


\begin{thebibliography}{70}
\ifx \bisbn   \undefined \def \bisbn  #1{ISBN #1}\fi
\ifx \binits  \undefined \def \binits#1{#1}\fi
\ifx \bauthor  \undefined \def \bauthor#1{#1}\fi
\ifx \batitle  \undefined \def \batitle#1{#1}\fi
\ifx \bjtitle  \undefined \def \bjtitle#1{#1}\fi
\ifx \bvolume  \undefined \def \bvolume#1{\textbf{#1}}\fi
\ifx \byear  \undefined \def \byear#1{#1}\fi
\ifx \bissue  \undefined \def \bissue#1{#1}\fi
\ifx \bfpage  \undefined \def \bfpage#1{#1}\fi
\ifx \blpage  \undefined \def \blpage #1{#1}\fi
\ifx \burl  \undefined \def \burl#1{\textsf{#1}}\fi
\ifx \doiurl  \undefined \def \doiurl#1{\url{https://doi.org/#1}}\fi
\ifx \betal  \undefined \def \betal{\textit{et al.}}\fi
\ifx \binstitute  \undefined \def \binstitute#1{#1}\fi
\ifx \binstitutionaled  \undefined \def \binstitutionaled#1{#1}\fi
\ifx \bctitle  \undefined \def \bctitle#1{#1}\fi
\ifx \beditor  \undefined \def \beditor#1{#1}\fi
\ifx \bpublisher  \undefined \def \bpublisher#1{#1}\fi
\ifx \bbtitle  \undefined \def \bbtitle#1{#1}\fi
\ifx \bedition  \undefined \def \bedition#1{#1}\fi
\ifx \bseriesno  \undefined \def \bseriesno#1{#1}\fi
\ifx \blocation  \undefined \def \blocation#1{#1}\fi
\ifx \bsertitle  \undefined \def \bsertitle#1{#1}\fi
\ifx \bsnm \undefined \def \bsnm#1{#1}\fi
\ifx \bsuffix \undefined \def \bsuffix#1{#1}\fi
\ifx \bparticle \undefined \def \bparticle#1{#1}\fi
\ifx \barticle \undefined \def \barticle#1{#1}\fi
\bibcommenthead
\ifx \bconfdate \undefined \def \bconfdate #1{#1}\fi
\ifx \botherref \undefined \def \botherref #1{#1}\fi
\ifx \url \undefined \def \url#1{\textsf{#1}}\fi
\ifx \bchapter \undefined \def \bchapter#1{#1}\fi
\ifx \bbook \undefined \def \bbook#1{#1}\fi
\ifx \bcomment \undefined \def \bcomment#1{#1}\fi
\ifx \oauthor \undefined \def \oauthor#1{#1}\fi
\ifx \citeauthoryear \undefined \def \citeauthoryear#1{#1}\fi
\ifx \endbibitem  \undefined \def \endbibitem {}\fi
\ifx \bconflocation  \undefined \def \bconflocation#1{#1}\fi
\ifx \arxivurl  \undefined \def \arxivurl#1{\textsf{#1}}\fi
\csname PreBibitemsHook\endcsname

\bibitem[\protect\citeauthoryear{{de Wit} et~al.}{2005}]{deWit2005}
\begin{barticle}
\bauthor{\bsnm{{de Wit}}, \binits{W.J.}},
\bauthor{\bsnm{{Testi}}, \binits{L.}},
\bauthor{\bsnm{{Palla}}, \binits{F.}},
\bauthor{\bsnm{{Zinnecker}}, \binits{H.}}:
\batitle{{The origin of massive O-type field stars: II. Field O stars as runaways}}.
\bjtitle{Astron. \& Astrophys.}
\bvolume{437}(\bissue{1}),
\bfpage{247}--\blpage{255}
(\byear{2005})
\end{barticle}
\endbibitem

\bibitem[\protect\citeauthoryear{{Fujii} and {Portegies Zwart}}{2011}]{Fujii2011}
\begin{barticle}
\bauthor{\bsnm{{Fujii}}, \binits{M.S.}},
\bauthor{\bsnm{{Portegies Zwart}}, \binits{S.}}:
\batitle{{The Origin of OB Runaway Stars}}.
\bjtitle{Science}
\bvolume{334}(\bissue{6061}),
\bfpage{1380}
(\byear{2011})
\end{barticle}
\endbibitem

\bibitem[\protect\citeauthoryear{{Gaia Collaboration et al.}}{2016}]{GaiaCollaboration2016}
\begin{barticle}
\bauthor{\bsnm{{Gaia Collaboration et al.}}}:
\batitle{{The Gaia mission}}.
\bjtitle{Astron. \& Astrophys}
\bvolume{595},
\bfpage{1}
(\byear{2016})
\end{barticle}
\endbibitem

\bibitem[\protect\citeauthoryear{{Gaia Collaboration et al.}}{2021}]{GaiaCollaboration2021}
\begin{barticle}
\bauthor{\bsnm{{Gaia Collaboration et al.}}}:
\batitle{{Gaia Early Data Release 3. Summary of the contents and survey properties}}.
\bjtitle{Astron. \& Astrophys.}
\bvolume{649},
\bfpage{1}
(\byear{2021})
\end{barticle}
\endbibitem

\bibitem[\protect\citeauthoryear{{Gaia Collaboration et al.}}{}]{GaiaCollaboration2023}
\begin{botherref}
\oauthor{\bsnm{{Gaia Collaboration et al.}}}:
{Gaia Data Release 3. Summary of the content and survey properties}.
Astron. \& Astrophys.
\textbf{674},
1
\end{botherref}
\endbibitem

\bibitem[\protect\citeauthoryear{{Banerjee} et~al.}{2012}]{Banerjee2012}
\begin{barticle}
\bauthor{\bsnm{{Banerjee}}, \binits{S.}},
\bauthor{\bsnm{{Kroupa}}, \binits{P.}},
\bauthor{\bsnm{{Oh}}, \binits{S.}}:
\batitle{{Runaway Massive Stars from R136: VFTS 682 is Very Likely a ``Slow Runaway''}}.
\bjtitle{Astrophys. J.}
\bvolume{746}(\bissue{1}),
\bfpage{15}
(\byear{2012})
\end{barticle}
\endbibitem

\bibitem[\protect\citeauthoryear{{Oh} et~al.}{2015}]{Oh2015}
\begin{barticle}
\bauthor{\bsnm{{Oh}}, \binits{S.}},
\bauthor{\bsnm{{Kroupa}}, \binits{P.}},
\bauthor{\bsnm{{Pflamm-Altenburg}}, \binits{J.}}:
\batitle{{Dependency of Dynamical Ejections of O Stars on the Masses of Very Young Star Clusters}}.
\bjtitle{Astrophys. J.}
\bvolume{805}(\bissue{2}),
\bfpage{92}
(\byear{2015})
\end{barticle}
\endbibitem

\bibitem[\protect\citeauthoryear{{Andersson} et~al.}{2020}]{Andersson2020}
\begin{barticle}
\bauthor{\bsnm{{Andersson}}, \binits{E.P.}},
\bauthor{\bsnm{{Agertz}}, \binits{O.}},
\bauthor{\bsnm{{Renaud}}, \binits{F.}}:
\batitle{{How runaway stars boost galactic outflows}}.
\bjtitle{Mon. Notices Royal Astron. Soc.}
\bvolume{494}(\bissue{3}),
\bfpage{3328}--\blpage{3341}
(\byear{2020})
\end{barticle}
\endbibitem

\bibitem[\protect\citeauthoryear{{Steinwandel} et~al.}{2023}]{Steinwandel2023}
\begin{barticle}
\bauthor{\bsnm{{Steinwandel}}, \binits{U.P.}},
\bauthor{\bsnm{{Bryan}}, \binits{G.L.}},
\bauthor{\bsnm{{Somerville}}, \binits{R.S.}},
\bauthor{\bsnm{{Hayward}}, \binits{C.C.}},
\bauthor{\bsnm{{Burkhart}}, \binits{B.}}:
\batitle{{On the impact of runaway stars on dwarf galaxies with resolved interstellar medium}}.
\bjtitle{Mon. Notices Royal Astron. Soc.}
\bvolume{526}(\bissue{1}),
\bfpage{1408}--\blpage{1427}
(\byear{2023})
\end{barticle}
\endbibitem

\bibitem[\protect\citeauthoryear{{Schneider, F. R. N. et al.}}{2018}]{Schneider2018}
\begin{barticle}
\bauthor{\bsnm{{Schneider, F. R. N. et al.}}}:
\batitle{{An excess of massive stars in the local 30 Doradus starburst}}.
\bjtitle{Science}
\bvolume{359}(\bissue{6371}),
\bfpage{69}--\blpage{71}
(\byear{2018})
\end{barticle}
\endbibitem

\bibitem[\protect\citeauthoryear{{Evans, C. J. et al.}}{2010}]{Evans2010}
\begin{barticle}
\bauthor{\bsnm{{Evans, C. J. et al.}}}:
\batitle{{A Massive Runaway Star from 30 Doradus}}.
\bjtitle{Astrophys. J. Lett.}
\bvolume{715}(\bissue{2}),
\bfpage{74}--\blpage{79}
(\byear{2010})
\end{barticle}
\endbibitem

\bibitem[\protect\citeauthoryear{{Lennon, D. J. et al.}}{2018}]{Lennon2018}
\begin{barticle}
\bauthor{\bsnm{{Lennon, D. J. et al.}}}:
\batitle{{Gaia DR2 reveals a very massive runaway star ejected from R136}}.
\bjtitle{Astron. \& Astrophys.}
\bvolume{619},
\bfpage{78}
(\byear{2018})
\end{barticle}
\endbibitem

\bibitem[\protect\citeauthoryear{{Sana, H. et al.}}{2022}]{Sana2022}
\begin{barticle}
\bauthor{\bsnm{{Sana, H. et al.}}}:
\batitle{{The VLT-FLAMES Tarantula Survey. Observational evidence for two distinct populations of massive runaway stars in 30 Doradus}}.
\bjtitle{Astron. \& Astrophys.}
\bvolume{668},
\bfpage{5}
(\byear{2022})
\end{barticle}
\endbibitem

\bibitem[\protect\citeauthoryear{{Oh} and {Kroupa}}{2016}]{Oh2016}
\begin{barticle}
\bauthor{\bsnm{{Oh}}, \binits{S.}},
\bauthor{\bsnm{{Kroupa}}, \binits{P.}}:
\batitle{{Dynamical ejections of massive stars from young star clusters under diverse initial conditions}}.
\bjtitle{Astron. \& Astrophys.}
\bvolume{590},
\bfpage{107}
(\byear{2016})
\end{barticle}
\endbibitem

\bibitem[\protect\citeauthoryear{{Ma{\'\i}z Apell{\'a}niz} et~al.}{2022}]{MaizApellaniz2022b_bermuda}
\begin{barticle}
\bauthor{\bsnm{{Ma{\'\i}z Apell{\'a}niz}}, \binits{J.}},
\bauthor{\bsnm{{Pantaleoni Gonz{\'a}lez}}, \binits{M.}},
\bauthor{\bsnm{{Barb{\'a}}}, \binits{R.H.}},
\bauthor{\bsnm{{Weiler}}, \binits{M.}}:
\batitle{{Escape from the Bermuda cluster: Orphanization by multiple stellar ejections}}.
\bjtitle{Astron. \& Astrophys.}
\bvolume{657},
\bfpage{72}
(\byear{2022})
\end{barticle}
\endbibitem

\bibitem[\protect\citeauthoryear{{Stoop, M. et al.}}{2024}]{Stoop2024}
\begin{barticle}
\bauthor{\bsnm{{Stoop, M. et al.}}}:
\batitle{{The early evolution of young massive clusters. II. The kinematic history of NGC 6618/M 17}}.
\bjtitle{Astron. \& Astrophys.}
\bvolume{681},
\bfpage{21}
(\byear{2024})
\end{barticle}
\endbibitem

\bibitem[\protect\citeauthoryear{{Brands, S. A. et al.}}{2022}]{Brands2022}
\begin{barticle}
\bauthor{\bsnm{{Brands, S. A. et al.}}}:
\batitle{{The R136 star cluster dissected with Hubble Space Telescope/STIS. III. The most massive stars and their clumped winds}}.
\bjtitle{Astron. \& Astrophys.}
\bvolume{663},
\bfpage{36}
(\byear{2022})
\end{barticle}
\endbibitem

\bibitem[\protect\citeauthoryear{{Crowther, P. A. et al.}}{2010}]{Crowther2010}
\begin{barticle}
\bauthor{\bsnm{{Crowther, P. A. et al.}}}:
\batitle{{The R136 star cluster hosts several stars whose individual masses greatly exceed the accepted 150M$_{solar}$ stellar mass limit}}.
\bjtitle{Mon. Notices Royal Astron. Soc.}
\bvolume{408}(\bissue{2}),
\bfpage{731}--\blpage{751}
(\byear{2010})
\end{barticle}
\endbibitem

\bibitem[\protect\citeauthoryear{{Bestenlehner, J. M. et al.}}{2020}]{Bestenlehner2020}
\begin{barticle}
\bauthor{\bsnm{{Bestenlehner, J. M. et al.}}}:
\batitle{{The R136 star cluster dissected with Hubble Space Telescope/STIS - II. Physical properties of the most massive stars in R136}}.
\bjtitle{Mon. Notices Royal Astron. Soc.}
\bvolume{499}(\bissue{2}),
\bfpage{1918}--\blpage{1936}
(\byear{2020})
\end{barticle}
\endbibitem

\bibitem[\protect\citeauthoryear{{Sabbi, E. et al.}}{2012}]{Sabbi2012}
\begin{barticle}
\bauthor{\bsnm{{Sabbi, E. et al.}}}:
\batitle{{A Double Cluster at the Core of 30 Doradus}}.
\bjtitle{Astrophys. J. Lett.}
\bvolume{754}(\bissue{2}),
\bfpage{37}
(\byear{2012})
\end{barticle}
\endbibitem

\bibitem[\protect\citeauthoryear{{Bestenlehner, J. M. et al.}}{2014}]{Bestenlehner2014}
\begin{barticle}
\bauthor{\bsnm{{Bestenlehner, J. M. et al.}}}:
\batitle{{The VLT-FLAMES Tarantula Survey. XVII. Physical and wind properties of massive stars at the top of the main sequence}}.
\bjtitle{Astron. \& Astrophys.}
\bvolume{570},
\bfpage{38}
(\byear{2014})
\end{barticle}
\endbibitem

\bibitem[\protect\citeauthoryear{{Crowther, P. A. et al.}}{2016}]{Crowther2016}
\begin{barticle}
\bauthor{\bsnm{{Crowther, P. A. et al.}}}:
\batitle{{The R136 star cluster dissected with Hubble Space Telescope/STIS. I. Far-ultraviolet spectroscopic census and the origin of He II {\ensuremath{\lambda}}1640 in young star clusters}}.
\bjtitle{Mon. Notices Royal Astron. Soc.}
\bvolume{458}(\bissue{1}),
\bfpage{624}--\blpage{659}
(\byear{2016})
\end{barticle}
\endbibitem

\bibitem[\protect\citeauthoryear{{Ram{\'\i}rez-Tannus, M. C. et al.}}{2021}]{RamirezTannus2021}
\begin{barticle}
\bauthor{\bsnm{{Ram{\'\i}rez-Tannus, M. C. et al.}}}:
\batitle{{A relation between the radial velocity dispersion of young clusters and their age. Evidence for hardening as the formation scenario of massive close binaries}}.
\bjtitle{Astron. \& Astrophys. Lett.}
\bvolume{645},
\bfpage{10}
(\byear{2021})
\end{barticle}
\endbibitem

\bibitem[\protect\citeauthoryear{{Salpeter}}{1955}]{Salpeter1955}
\begin{barticle}
\bauthor{\bsnm{{Salpeter}}, \binits{E.E.}}:
\batitle{{The Luminosity Function and Stellar Evolution.}}
\bjtitle{Astrophys. J.}
\bvolume{121},
\bfpage{161}
(\byear{1955})
\end{barticle}
\endbibitem

\bibitem[\protect\citeauthoryear{{Doran, E. I. et al.}}{2013}]{Doran2013}
\begin{barticle}
\bauthor{\bsnm{{Doran, E. I. et al.}}}:
\batitle{{The VLT-FLAMES Tarantula Survey. XI. A census of the hot luminous stars and their feedback in 30 Doradus}}.
\bjtitle{Astron. \& Astrophys.}
\bvolume{558},
\bfpage{134}
(\byear{2013})
\end{barticle}
\endbibitem

\bibitem[\protect\citeauthoryear{{Barkana} and {Loeb}}{2001}]{Barkana2001}
\begin{barticle}
\bauthor{\bsnm{{Barkana}}, \binits{R.}},
\bauthor{\bsnm{{Loeb}}, \binits{A.}}
\bjtitle{Phys. Rep.}
\bvolume{349}(\bissue{2}),
\bfpage{125}--\blpage{238}
(\byear{2001})
\end{barticle}
\endbibitem

\bibitem[\protect\citeauthoryear{{Wise} and {Cen}}{2009}]{Wise2009}
\begin{barticle}
\bauthor{\bsnm{{Wise}}, \binits{J.H.}},
\bauthor{\bsnm{{Cen}}, \binits{R.}}:
\batitle{{Ionizing Photon Escape Fractions From High-Redshift Dwarf Galaxies}}.
\bjtitle{Astrophys. J.}
\bvolume{693}(\bissue{1}),
\bfpage{984}--\blpage{999}
(\byear{2009})
\end{barticle}
\endbibitem

\bibitem[\protect\citeauthoryear{{Atek, H. et al.}}{2024}]{Atek2023}
\begin{barticle}
\bauthor{\bsnm{{Atek, H. et al.}}}:
\batitle{{Most of the photons that reionized the Universe came from dwarf galaxies}}.
\bjtitle{Nature}
\bvolume{626}(\bissue{8001}),
\bfpage{975}--\blpage{978}
(\byear{2024})
\end{barticle}
\endbibitem

\bibitem[\protect\citeauthoryear{{Razoumov} and {Sommer-Larsen}}{2007}]{Razoumov2007}
\begin{barticle}
\bauthor{\bsnm{{Razoumov}}, \binits{A.O.}},
\bauthor{\bsnm{{Sommer-Larsen}}, \binits{J.}}:
\batitle{{Modeling Lyman Continuum Emission from Young Galaxies}}.
\bjtitle{Astrophys. J.}
\bvolume{668}(\bissue{2}),
\bfpage{674}--\blpage{681}
(\byear{2007})
\doiurl{10.1086/521041}
\end{barticle}
\endbibitem

\bibitem[\protect\citeauthoryear{{Conroy} and {Kratter}}{2012}]{Conroy2012}
\begin{barticle}
\bauthor{\bsnm{{Conroy}}, \binits{C.}},
\bauthor{\bsnm{{Kratter}}, \binits{K.M.}}:
\batitle{{Runaway Stars and the Escape of Ionizing Radiation from High-redshift Galaxies}}.
\bjtitle{Astrophys. J.}
\bvolume{755}(\bissue{2}),
\bfpage{123}
(\byear{2012})
\end{barticle}
\endbibitem

\bibitem[\protect\citeauthoryear{{Pietrzy{\'n}ski, G. et al.}}{2019}]{Pietrzynski2019}
\begin{barticle}
\bauthor{\bsnm{{Pietrzy{\'n}ski, G. et al.}}}:
\batitle{{A distance to the Large Magellanic Cloud that is precise to one per cent}}.
\bjtitle{Nature}
\bvolume{567}(\bissue{7747}),
\bfpage{200}--\blpage{203}
(\byear{2019})
\end{barticle}
\endbibitem

\bibitem[\protect\citeauthoryear{{Lindegren, L. et al.}}{2021}]{Lindegren2021}
\begin{barticle}
\bauthor{\bsnm{{Lindegren, L. et al.}}}:
\batitle{{Gaia Early Data Release 3. Parallax bias versus magnitude, colour, and position}}.
\bjtitle{Astron. \& Astrophys.}
\bvolume{649},
\bfpage{4}
(\byear{2021})
\end{barticle}
\endbibitem

\bibitem[\protect\citeauthoryear{Lindegren}{2018}]{Lindegren2018}
\begin{botherref}
\oauthor{\bsnm{Lindegren}, \binits{L.}}:
{R}e-normalising the astrometric chi-square in {G}aia {D}{R}2.
GAIA-C3-TN-LU-LL-124
(2018).
\url{http://www.rssd.esa.int/doc_fetch.php?id=3757412}
\end{botherref}
\endbibitem

\bibitem[\protect\citeauthoryear{{Pecaut} and {Mamajek}}{2013}]{Pecaut2013}
\begin{barticle}
\bauthor{\bsnm{{Pecaut}}, \binits{M.J.}},
\bauthor{\bsnm{{Mamajek}}, \binits{E.E.}}:
\batitle{{Intrinsic Colors, Temperatures, and Bolometric Corrections of Pre-main-sequence Stars}}.
\bjtitle{Astrophys. J.s}
\bvolume{208}(\bissue{1}),
\bfpage{9}
(\byear{2013})
\end{barticle}
\endbibitem

\bibitem[\protect\citeauthoryear{{H{\'e}nault-Brunet, V. et al.}}{2012}]{HenaultBrunet2012}
\begin{barticle}
\bauthor{\bsnm{{H{\'e}nault-Brunet, V. et al.}}}:
\batitle{{The VLT-FLAMES Tarantula Survey. VII. A low velocity dispersion for the young massive cluster R136}}.
\bjtitle{Astron. \& Astrophys.}
\bvolume{546},
\bfpage{73}
(\byear{2012})
\end{barticle}
\endbibitem

\bibitem[\protect\citeauthoryear{{Skrutskie, M. F. et al.}}{2006}]{Skrutskie2006}
\begin{barticle}
\bauthor{\bsnm{{Skrutskie, M. F. et al.}}}:
\batitle{{The Two Micron All Sky Survey (2MASS)}}.
\bjtitle{Astron. J}
\bvolume{131}(\bissue{2}),
\bfpage{1163}--\blpage{1183}
(\byear{2006})
\end{barticle}
\endbibitem

\bibitem[\protect\citeauthoryear{{Bestenlehner, J. M. et al.}}{2011}]{Bestenlehner2011}
\begin{barticle}
\bauthor{\bsnm{{Bestenlehner, J. M. et al.}}}:
\batitle{{The VLT-FLAMES Tarantula Survey. III. A very massive star in apparent isolation from the massive cluster R136}}.
\bjtitle{Astron. \& Astrophys.}
\bvolume{530},
\bfpage{14}
(\byear{2011})
\end{barticle}
\endbibitem

\bibitem[\protect\citeauthoryear{{Renzo, M. et al.}}{2019}]{Renzo2019vfts682}
\begin{barticle}
\bauthor{\bsnm{{Renzo, M. et al.}}}:
\batitle{{Space astrometry of the very massive {\ensuremath{\sim}}150 M$_{{\ensuremath{\odot}}}$ candidate runaway star VFTS682}}.
\bjtitle{Mon. Notices Royal Astron. Soc.}
\bvolume{482}(\bissue{1}),
\bfpage{102}--\blpage{106}
(\byear{2019})
\end{barticle}
\endbibitem

\bibitem[\protect\citeauthoryear{{Carretero-Castrillo} et~al.}{2023}]{CarreteroCastrillo2023}
\begin{barticle}
\bauthor{\bsnm{{Carretero-Castrillo}}, \binits{M.}},
\bauthor{\bsnm{{Rib{\'o}}}, \binits{M.}},
\bauthor{\bsnm{{Paredes}}, \binits{J.M.}}:
\batitle{{Galactic runaway O and Be stars found using Gaia DR3}}.
\bjtitle{Astron. \& Astrophys.}
\bvolume{679},
\bfpage{109}
(\byear{2023})
\end{barticle}
\endbibitem

\bibitem[\protect\citeauthoryear{{Portegies Zwart} et~al.}{2010}]{PortegiesZwart2010}
\begin{barticle}
\bauthor{\bsnm{{Portegies Zwart}}, \binits{S.F.}},
\bauthor{\bsnm{{McMillan}}, \binits{S.L.W.}},
\bauthor{\bsnm{{Gieles}}, \binits{M.}}:
\batitle{{Young Massive Star Clusters}}.
\bjtitle{Annu. Rev. Astron. Astrophys.}
\bvolume{48},
\bfpage{431}--\blpage{493}
(\byear{2010})
\end{barticle}
\endbibitem

\bibitem[\protect\citeauthoryear{{Brands, S. A. et al.}}{2023}]{Brands2023}
\begin{barticle}
\bauthor{\bsnm{{Brands, S. A. et al.}}}:
\batitle{{Extinction towards the cluster R136 in the Large Magellanic Cloud. An extinction law from the near-infrared to the ultraviolet}}.
\bjtitle{Astron. \& Astrophys.}
\bvolume{673},
\bfpage{132}
(\byear{2023})
\end{barticle}
\endbibitem

\bibitem[\protect\citeauthoryear{{Tehrani, K. A. et al.}}{2019}]{Tehrani2019}
\begin{barticle}
\bauthor{\bsnm{{Tehrani, K. A. et al.}}}:
\batitle{{Weighing Melnick 34: the most massive binary system known}}.
\bjtitle{Mon. Notices Royal Astron. Soc.}
\bvolume{484}(\bissue{2}),
\bfpage{2692}--\blpage{2710}
(\byear{2019})
\end{barticle}
\endbibitem

\bibitem[\protect\citeauthoryear{{Schnurr} et~al.}{2008}]{Schnurr2008}
\begin{barticle}
\bauthor{\bsnm{{Schnurr}}, \binits{O.}},
\bauthor{\bsnm{{Moffat}}, \binits{A.F.J.}},
\bauthor{\bsnm{{St-Louis}}, \binits{N.}},
\bauthor{\bsnm{{Morrell}}, \binits{N.I.}},
\bauthor{\bsnm{{Guerrero}}, \binits{M.A.}}:
\batitle{{A spectroscopic survey of WNL stars in the Large Magellanic Cloud: general properties and binary status}}.
\bjtitle{Mon. Notices Royal Astron. Soc.}
\bvolume{389}(\bissue{2}),
\bfpage{806}--\blpage{828}
(\byear{2008})
\end{barticle}
\endbibitem

\bibitem[\protect\citeauthoryear{{Shenar, T. et al.}}{2019}]{Shenar2019}
\begin{barticle}
\bauthor{\bsnm{{Shenar, T. et al.}}}:
\batitle{{The Wolf-Rayet binaries of the nitrogen sequence in the Large Magellanic Cloud. Spectroscopy, orbital analysis, formation, and evolution}}.
\bjtitle{Astron. \& Astrophys.}
\bvolume{627},
\bfpage{151}
(\byear{2019})
\end{barticle}
\endbibitem

\bibitem[\protect\citeauthoryear{{Castro, N. et al.}}{2018}]{Castro2018}
\begin{barticle}
\bauthor{\bsnm{{Castro, N. et al.}}}:
\batitle{{Mapping the core of the Tarantula Nebula with VLT-MUSE. I. Spectral and nebular content around R136}}.
\bjtitle{Astron. \& Astrophys.}
\bvolume{614},
\bfpage{147}
(\byear{2018})
\end{barticle}
\endbibitem

\bibitem[\protect\citeauthoryear{{Sana, H. et al.}}{2013}]{Sana2013}
\begin{barticle}
\bauthor{\bsnm{{Sana, H. et al.}}}:
\batitle{{The VLT-FLAMES Tarantula Survey. VIII. Multiplicity properties of the O-type star population}}.
\bjtitle{Astron. \& Astrophys.}
\bvolume{550},
\bfpage{107}
(\byear{2013})
\end{barticle}
\endbibitem

\bibitem[\protect\citeauthoryear{{Mahy, L. et al.}}{2020}]{Mahy2020}
\begin{barticle}
\bauthor{\bsnm{{Mahy, L. et al.}}}:
\batitle{{The Tarantula Massive Binary Monitoring. III. Atmosphere analysis of double-lined spectroscopic systems}}.
\bjtitle{Astron. \& Astrophys.}
\bvolume{634},
\bfpage{118}
(\byear{2020})
\end{barticle}
\endbibitem

\bibitem[\protect\citeauthoryear{{Shenar, T. et al.}}{2021}]{Shenar2021}
\begin{barticle}
\bauthor{\bsnm{{Shenar, T. et al.}}}:
\batitle{{The Tarantula Massive Binary Monitoring. V. R 144: a wind-eclipsing binary with a total mass {\ensuremath{\gtrsim}}140 M$_{{\ensuremath{\odot}}}$}}.
\bjtitle{Astron. \& Astrophys.}
\bvolume{650},
\bfpage{147}
(\byear{2021})
\end{barticle}
\endbibitem

\bibitem[\protect\citeauthoryear{{Sab{\'\i}n-Sanjuli{\'a}n, C. et al.}}{2017}]{SabinSanjulian2017}
\begin{barticle}
\bauthor{\bsnm{{Sab{\'\i}n-Sanjuli{\'a}n, C. et al.}}}:
\batitle{{The VLT-FLAMES Tarantula Survey. XXVI. Properties of the O-dwarf population in 30 Doradus}}.
\bjtitle{Astron. \& Astrophys.}
\bvolume{601},
\bfpage{79}
(\byear{2017})
\end{barticle}
\endbibitem

\bibitem[\protect\citeauthoryear{{McEvoy, C. M. et al.}}{2015}]{McEvoy2015}
\begin{barticle}
\bauthor{\bsnm{{McEvoy, C. M. et al.}}}:
\batitle{{The VLT-FLAMES Tarantula Survey. XIX. B-type supergiants: Atmospheric parameters and nitrogen abundances to investigate the role of binarity and the width of the main sequence}}.
\bjtitle{Astron. \& Astrophys.}
\bvolume{575},
\bfpage{70}
(\byear{2015})
\end{barticle}
\endbibitem

\bibitem[\protect\citeauthoryear{{Evans} et~al.}{2015}]{Evans2015b}
\begin{barticle}
\bauthor{\bsnm{{Evans}}, \binits{C.J.}},
\bauthor{\bsnm{{van Loon}}, \binits{J.T.}},
\bauthor{\bsnm{{Hainich}}, \binits{R.}},
\bauthor{\bsnm{{Bailey}}, \binits{M.}}:
\batitle{{2dF-AAOmega spectroscopy of massive stars in the Magellanic Clouds. The north-eastern region of the Large Magellanic Cloud}}.
\bjtitle{Astron. \& Astrophys.}
\bvolume{584},
\bfpage{5}
(\byear{2015})
\end{barticle}
\endbibitem

\bibitem[\protect\citeauthoryear{{Walborn} et~al.}{1995}]{Walborn1995}
\begin{barticle}
\bauthor{\bsnm{{Walborn}}, \binits{N.R.}},
\bauthor{\bsnm{{Lennon}}, \binits{D.J.}},
\bauthor{\bsnm{{Haser}}, \binits{S.M.}},
\bauthor{\bsnm{{Kudritzki}}, \binits{R.-P.}},
\bauthor{\bsnm{{Voels}}, \binits{S.A.}}:
\batitle{{The Physics of Massive OB Stars in Different Parent Galaxies. I. Ultraviolet and Optical Spectral Morphology in the Magellanic Clouds}}.
\bjtitle{Publ. Astron. Soc. Pac.}
\bvolume{107},
\bfpage{104}
(\byear{1995})
\end{barticle}
\endbibitem

\bibitem[\protect\citeauthoryear{{Walborn, N. R. et al.}}{2004}]{Walborn2004}
\begin{barticle}
\bauthor{\bsnm{{Walborn, N. R. et al.}}}:
\batitle{{A CNO Dichotomy among O2 Giant Spectra in the Magellanic Clouds}}.
\bjtitle{Astrophys. J.}
\bvolume{608}(\bissue{2}),
\bfpage{1028}--\blpage{1038}
(\byear{2004})
\end{barticle}
\endbibitem

\bibitem[\protect\citeauthoryear{{Martins} et~al.}{2005}]{Martins2005}
\begin{barticle}
\bauthor{\bsnm{{Martins}}, \binits{F.}},
\bauthor{\bsnm{{Schaerer}}, \binits{D.}},
\bauthor{\bsnm{{Hillier}}, \binits{D.J.}}:
\batitle{{A new calibration of stellar parameters of Galactic O stars}}.
\bjtitle{Astron. \& Astrophys.}
\bvolume{436}(\bissue{3}),
\bfpage{1049}--\blpage{1065}
(\byear{2005})
\end{barticle}
\endbibitem

\bibitem[\protect\citeauthoryear{{Gvaramadze} et~al.}{2010}]{Gvaramadze2010}
\begin{barticle}
\bauthor{\bsnm{{Gvaramadze}}, \binits{V.V.}},
\bauthor{\bsnm{{Kroupa}}, \binits{P.}},
\bauthor{\bsnm{{Pflamm-Altenburg}}, \binits{J.}}:
\batitle{{Massive runaway stars in the Large Magellanic Cloud}}.
\bjtitle{Astron. \& Astrophys.}
\bvolume{519},
\bfpage{33}
(\byear{2010})
\end{barticle}
\endbibitem

\bibitem[\protect\citeauthoryear{Stoppa et~al.}{2023}]{Stoppa2023}
\begin{barticle}
\bauthor{\bsnm{Stoppa}, \binits{F.}},
\bauthor{\bsnm{Cator}, \binits{E.}},
\bauthor{\bsnm{Nelemans}, \binits{G.}}:
\batitle{{Consistency tests for comparing astrophysical models and observations}}.
\bjtitle{Monthly Notices of the Royal Astronomical Society}
\bvolume{524}(\bissue{1}),
\bfpage{1061}--\blpage{1074}
(\byear{2023})
\end{barticle}
\endbibitem

\bibitem[\protect\citeauthoryear{{Perets} and {{\v{S}}ubr}}{2012}]{Perets2012}
\begin{barticle}
\bauthor{\bsnm{{Perets}}, \binits{H.B.}},
\bauthor{\bsnm{{{\v{S}}ubr}}, \binits{L.}}:
\batitle{{The Properties of Dynamically Ejected Runaway and Hyper-runaway Stars}}.
\bjtitle{Astrophys. J.}
\bvolume{751}(\bissue{2}),
\bfpage{133}
(\byear{2012})
\end{barticle}
\endbibitem

\bibitem[\protect\citeauthoryear{MacQueen}{1967}]{MacQueen1967}
\begin{botherref}
\oauthor{\bsnm{MacQueen}, \binits{J.}}:
Some methods for classification and analysis of multivariate observations
(1967)
\end{botherref}
\endbibitem

\bibitem[\protect\citeauthoryear{{Evans, C. J. et al.}}{2015}]{Evans2015a}
\begin{barticle}
\bauthor{\bsnm{{Evans, C. J. et al.}}}:
\batitle{{The VLT-FLAMES Tarantula Survey. XVIII. Classifications and radial velocities of the B-type stars}}.
\bjtitle{Astron. \& Astrophys.}
\bvolume{574},
\bfpage{13}
(\byear{2015})
\end{barticle}
\endbibitem

\bibitem[\protect\citeauthoryear{{Castro, N. et al.}}{2021}]{Castro2021}
\begin{barticle}
\bauthor{\bsnm{{Castro, N. et al.}}}:
\batitle{{Mapping the core of the Tarantula Nebula with VLT-MUSE. II. The spectroscopic Hertzsprung-Russell diagram of OB stars in NGC 2070}}.
\bjtitle{Astron. \& Astrophys.}
\bvolume{648},
\bfpage{65}
(\byear{2021})
\end{barticle}
\endbibitem

\bibitem[\protect\citeauthoryear{{Walborn, N. R. et al.}}{2014}]{Walborn2014}
\begin{barticle}
\bauthor{\bsnm{{Walborn, N. R. et al.}}}:
\batitle{{The VLT-FLAMES Tarantula Survey. XIV. The O-type stellar content of 30 Doradus}}.
\bjtitle{Astron. \& Astrophys.}
\bvolume{564},
\bfpage{40}
(\byear{2014})
\end{barticle}
\endbibitem

\bibitem[\protect\citeauthoryear{{van Gelder, M.~L. et al.}}{2020}]{vanGelder2020}
\begin{barticle}
\bauthor{\bsnm{{van Gelder, M.~L. et al.}}}:
\batitle{{VLT/X-shooter spectroscopy of massive young stellar objects in the 30 Doradus region of the Large Magellanic Cloud}}.
\bjtitle{Astron. \& Astrophys.}
\bvolume{636},
\bfpage{54}
(\byear{2020})
\end{barticle}
\endbibitem

\bibitem[\protect\citeauthoryear{{Ram{\'\i}rez-Agudelo, O. H. et al.}}{2017}]{RamirezAgudelo2017}
\begin{barticle}
\bauthor{\bsnm{{Ram{\'\i}rez-Agudelo, O. H. et al.}}}:
\batitle{{The VLT-FLAMES Tarantula Survey . XXIV. Stellar properties of the O-type giants and supergiants in 30 Doradus}}.
\bjtitle{Astron. \& Astrophys.}
\bvolume{600},
\bfpage{81}
(\byear{2017})
\end{barticle}
\endbibitem

\bibitem[\protect\citeauthoryear{{Evans, C. J. et al.}}{2011}]{Evans2011}
\begin{barticle}
\bauthor{\bsnm{{Evans, C. J. et al.}}}:
\batitle{{The VLT-FLAMES Tarantula Survey. I. Introduction and observational overview}}.
\bjtitle{Astron. \& Astrophys.}
\bvolume{530},
\bfpage{108}
(\byear{2011})
\end{barticle}
\endbibitem

\bibitem[\protect\citeauthoryear{{Shenar, T. et al.}}{2022}]{Shenar2022}
\begin{barticle}
\bauthor{\bsnm{{Shenar, T. et al.}}}:
\batitle{{The Tarantula Massive Binary Monitoring. VI. Characterisation of hidden companions in 51 single-lined O-type binaries: A flat mass-ratio distribution and black-hole binary candidates}}.
\bjtitle{Astron. \& Astrophys.}
\bvolume{665},
\bfpage{148}
(\byear{2022})
\end{barticle}
\endbibitem

\bibitem[\protect\citeauthoryear{{Dufton, P. L. et al.}}{2013}]{Dufton2013}
\begin{barticle}
\bauthor{\bsnm{{Dufton, P. L. et al.}}}:
\batitle{{The VLT-FLAMES Tarantula Survey. X. Evidence for a bimodal distribution of rotational velocities for the single early B-type stars}}.
\bjtitle{Astron. \& Astrophys.}
\bvolume{550},
\bfpage{109}
(\byear{2013})
\end{barticle}
\endbibitem

\bibitem[\protect\citeauthoryear{{Foellmi} et~al.}{2003}]{Foellmi2003}
\begin{barticle}
\bauthor{\bsnm{{Foellmi}}, \binits{C.}},
\bauthor{\bsnm{{Moffat}}, \binits{A.F.J.}},
\bauthor{\bsnm{{Guerrero}}, \binits{M.A.}}:
\batitle{{Wolf-Rayet binaries in the Magellanic Clouds and implications for massive-star evolution - II. Large Magellanic Cloud}}.
\bjtitle{Mon. Notices Royal Astron. Soc.}
\bvolume{338}(\bissue{4}),
\bfpage{1025}--\blpage{1056}
(\byear{2003})
\end{barticle}
\endbibitem

\bibitem[\protect\citeauthoryear{{Villase{\~n}or, J. I. et al.}}{2021}]{Villasenor2021}
\begin{barticle}
\bauthor{\bsnm{{Villase{\~n}or, J. I. et al.}}}:
\batitle{{The B-type binaries characterization programme I. Orbital solutions for the 30 Doradus population}}.
\bjtitle{Mon. Notices Royal Astron. Soc.}
\bvolume{507}(\bissue{4}),
\bfpage{5348}--\blpage{5375}
(\byear{2021})
\end{barticle}
\endbibitem

\bibitem[\protect\citeauthoryear{{Dufton, P. L. et al.}}{2018}]{Dufton2018}
\begin{barticle}
\bauthor{\bsnm{{Dufton, P. L. et al.}}}:
\batitle{{The VLT-FLAMES Tarantula Survey. XXVIII. Nitrogen abundances for apparently single dwarf and giant B-type stars with small projected rotational velocities}}.
\bjtitle{Astron. \& Astrophys.}
\bvolume{615},
\bfpage{101}
(\byear{2018})
\end{barticle}
\endbibitem

\bibitem[\protect\citeauthoryear{{Kamath} et~al.}{2015}]{Kamath2015}
\begin{barticle}
\bauthor{\bsnm{{Kamath}}, \binits{D.}},
\bauthor{\bsnm{{Wood}}, \binits{P.R.}},
\bauthor{\bsnm{{Van Winckel}}, \binits{H.}}:
\batitle{{Optically visible post-AGB stars, post-RGB stars and young stellar objects in the Large Magellanic Cloud}}.
\bjtitle{Mon. Notices Royal Astron. Soc.}
\bvolume{454}(\bissue{2}),
\bfpage{1468}--\blpage{1502}
(\byear{2015})
\end{barticle}
\endbibitem

\end{thebibliography}

\section*{Acknowledgements}
\label{sec:Acknowledgements}
We thank Frank Backs, Omar Ould-Boukattine, Jason Hessels, Dion Linssen, and Mark Snelders for their help and discussions. MS acknowledges support from NOVA. The research leading to these results has received funding from the European Research Council (ERC) under the European Union's Horizon 2020 research and innovation programme (grant agreement numbers 772225: MULTIPLES). This work has made use of data from the European Space Agency (ESA) mission
{\it Gaia} (\url{https://www.cosmos.esa.int/gaia}), processed by the {\it Gaia}
Data Processing and Analysis Consortium (DPAC,
\url{https://www.cosmos.esa.int/web/gaia/dpac/consortium}). Funding for the DPAC
has been provided by national institutions, in particular the institutions
participating in the {\it Gaia} Multilateral Agreement. This publication makes use of data products from the Two Micron All Sky Survey, which is a joint project of the University of Massachusetts and the Infrared Processing and Analysis Center/California Institute of Technology, funded by the National Aeronautics and Space Administration and the National Science Foundation.

\section*{Author contributions}
MS led the runaway search, data analysis, produced all figures and tables in this study, and is the main contributor to the text. AdK and LK contributed to the scientific interpretation, context and text. SB contributed to the results, implications and text. SPZ and HS contributed to the scientific interpretation, implications and context. FS contributed to the statistical interpretation. All other authors contributed towards the discussion and provided feedback on the text.

\section*{Declarations}
\label{sec:Declarations}
All authors have no competing interests.

\section*{Supplementary Information}
Supplementary Information is available for this paper.

\section*{Author correspondence}
Correspondence and requests for materials should be addressed to M. Stoop.

\subsubsection*{Reprints and permissions information is available at \url{www.nature.com/reprints.}}

\clearpage
\begin{appendices}
\section*{Extended Data}

\renewcommand{\figurename}{Extended Data Fig.}
\begin{figure}[h]
\centering
\includegraphics[width=0.99\linewidth]{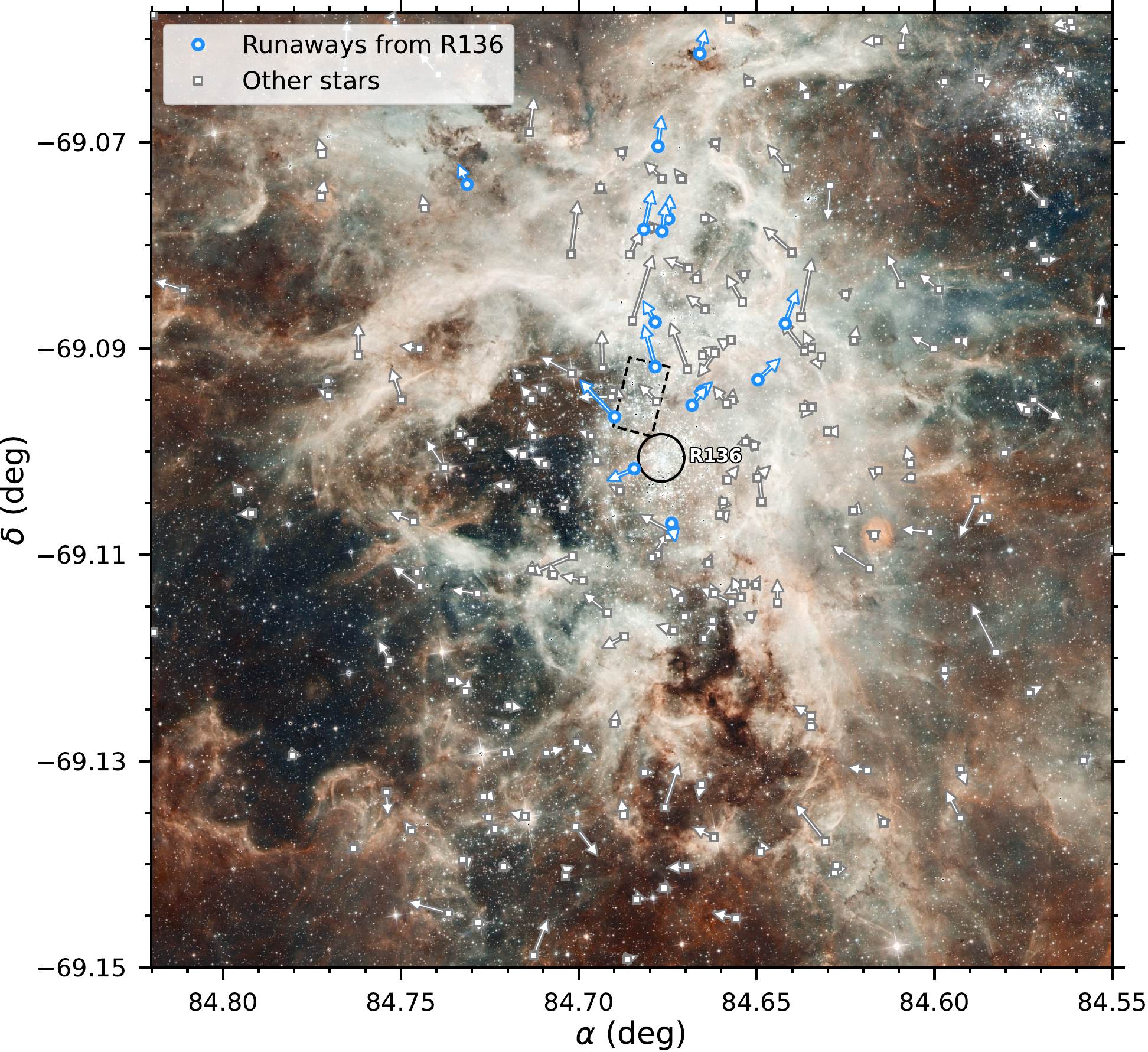}
\caption{\textbf{$\vert$ Proper motion of stars relative to R136 in the field in and around R136 with reliable astrometry}. The blue stars represent the runaways coming from R136 found in this work, while the grey stars do not originate in R136. The stars with reliable astrometry are defined in Section~\ref{sec:methods_gaia_filters}. The proper motion of R136 found in Section~\ref{sec:methods_r136_pm} has been subtracted and is $\mu_{\alpha^{*},\rm{R136}}$ = 1.654 mas yr$^{-1}$ and $\mu_{\delta,\rm{R136}}$ = 0.573 mas yr$^{-1}$. The black circle is centred on R136 and has a radius of 2.0 pc. The dashed black rectangle depicts the region used to derive the colour-magnitude-diagram of the north-east cluster in \cite{Sabbi2012}. The background image is taken by the \textit{Hubble Space Telescope} (\textit{HST}) and European Southern Observatory (ESO) 2.2m telescope (NASA, ESA \& Lennon et al.; 2012).}
\label{fig:r136_good_astrometry}
\end{figure}

\renewcommand{\figurename}{Extended Data Fig.}
\begin{figure}
\centering
\includegraphics[width=0.99\linewidth]{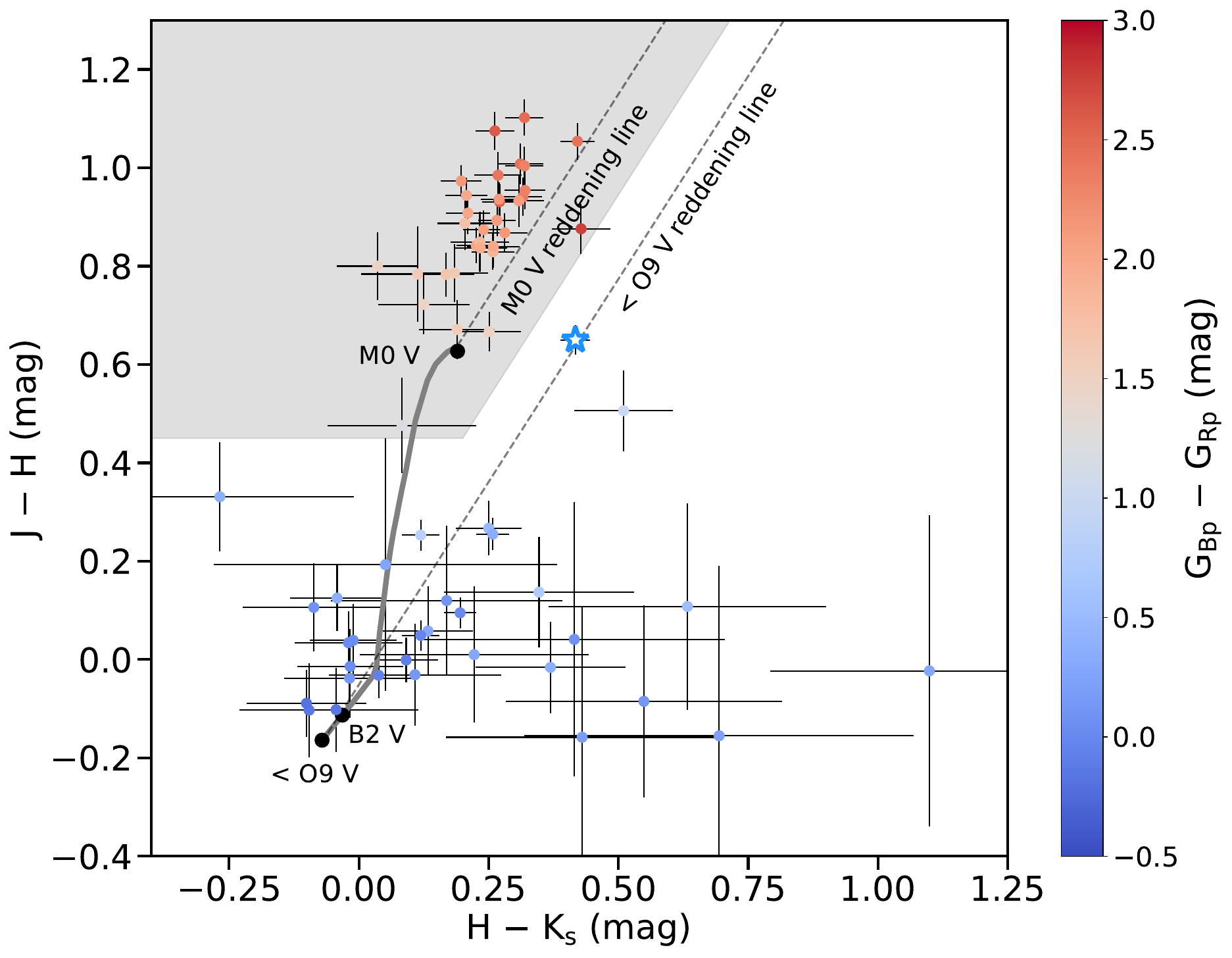}
\caption{\textbf{$\vert$ Near-infrared (\textit{2MASS}) colour-colour diagram of the runaway candidates}. They are coloured according to their \textit{Gaia} colour G$_{\rm{Bp}}$ $-$ G$_{\rm{Rp}}$. The location of an O-type ($<$ O9 V), B2 V, and M0 V star on the zero-age-main-sequence in this diagram are indicated with the black circles \citep{Pecaut2013}. The location of main-sequence stars with spectral types between O9 V and M0 V are represented by the grey curve. The reddening lines for the O9 and M0 dwarf star are given with the black dashed lines. The reddened WN5h star VFTS 682 is shown with the open blue star. Stars in the grey shaded region in the upper left corner are excluded from the final runaway sample as they are consistent with late-type stars. The uncertainty on the data is given as a 1$\sigma$ confidence interval.}
\label{fig:runaways_jh_hk}
\end{figure}

\renewcommand{\figurename}{Extended Data Fig.}
\begin{figure}
\centering
\includegraphics[width=0.99\linewidth]{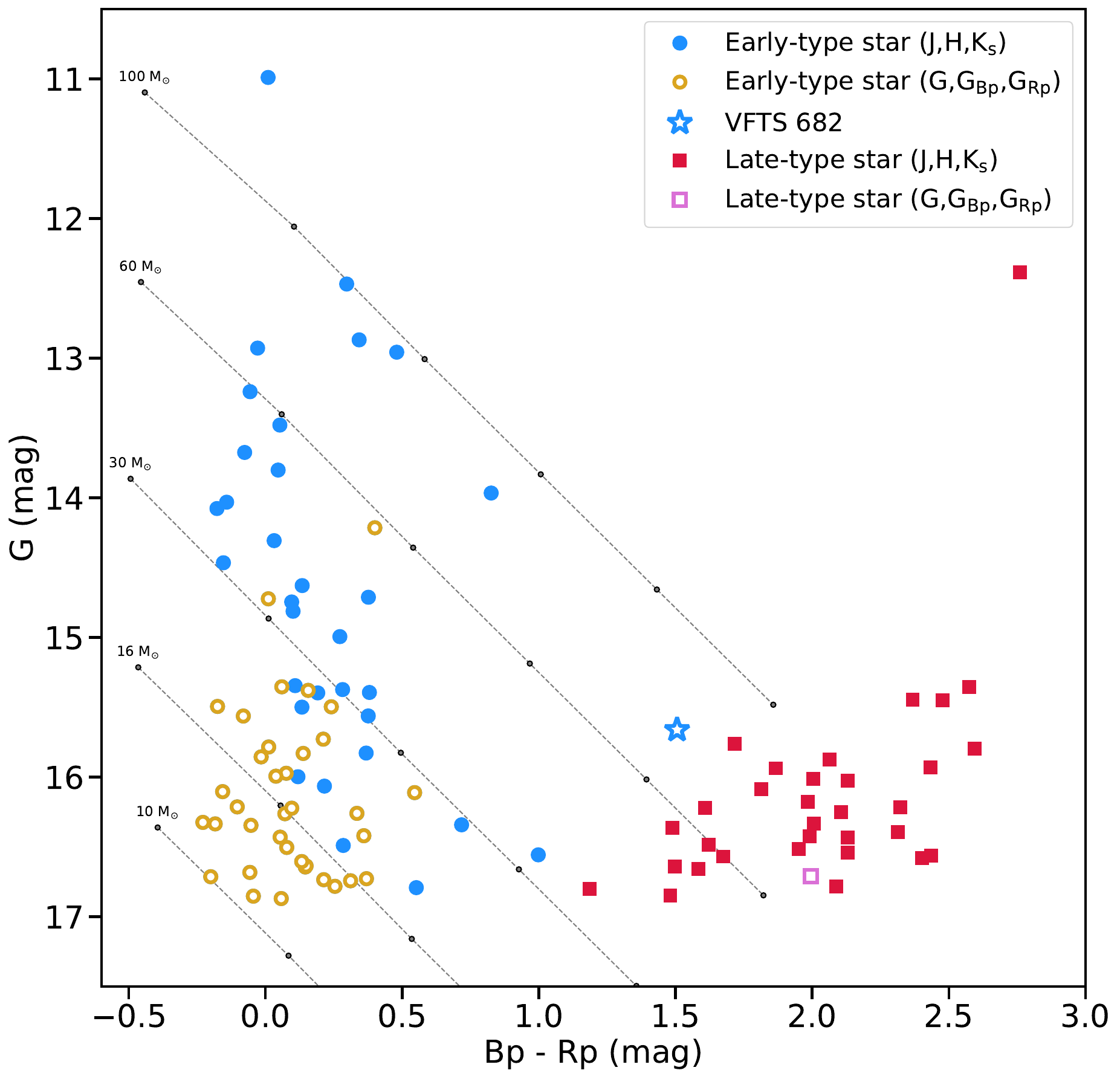}
\caption{\textbf{$\vert$ \textit{Gaia} colour magnitude diagram (CMD) of the runaway candidates}. Stars with blue circles and red squares were included and excluded, respectively, based on Figure~\ref{fig:runaways_jh_hk}. Parsec reddening lines are shown for a 10, 16, 30, 60, and 100 M$_{\odot}$ star with the grey dots denoting $A_{\rm{V}}$ equal to 0.0 to 5.0 mag in steps of 1.0 mag ($R_{\rm{V}}$ = 3.1) for an age = 1.8 Myr. The relatively reddened WN5h star VFTS 682 is shown with the open blue star. The open yellow circles are included in the final runaway sample on the basis of their relatively blue colour. The open purple square is excluded based on its relatively red colour. The star located in the upper right is the red supergiant MH 18 that could be a massive runaway star.}
\label{fig:runaways_g_bprp}
\end{figure}

\renewcommand{\figurename}{Extended Data Fig.}
\begin{figure*}
\centering
\includegraphics[width=0.99\linewidth]{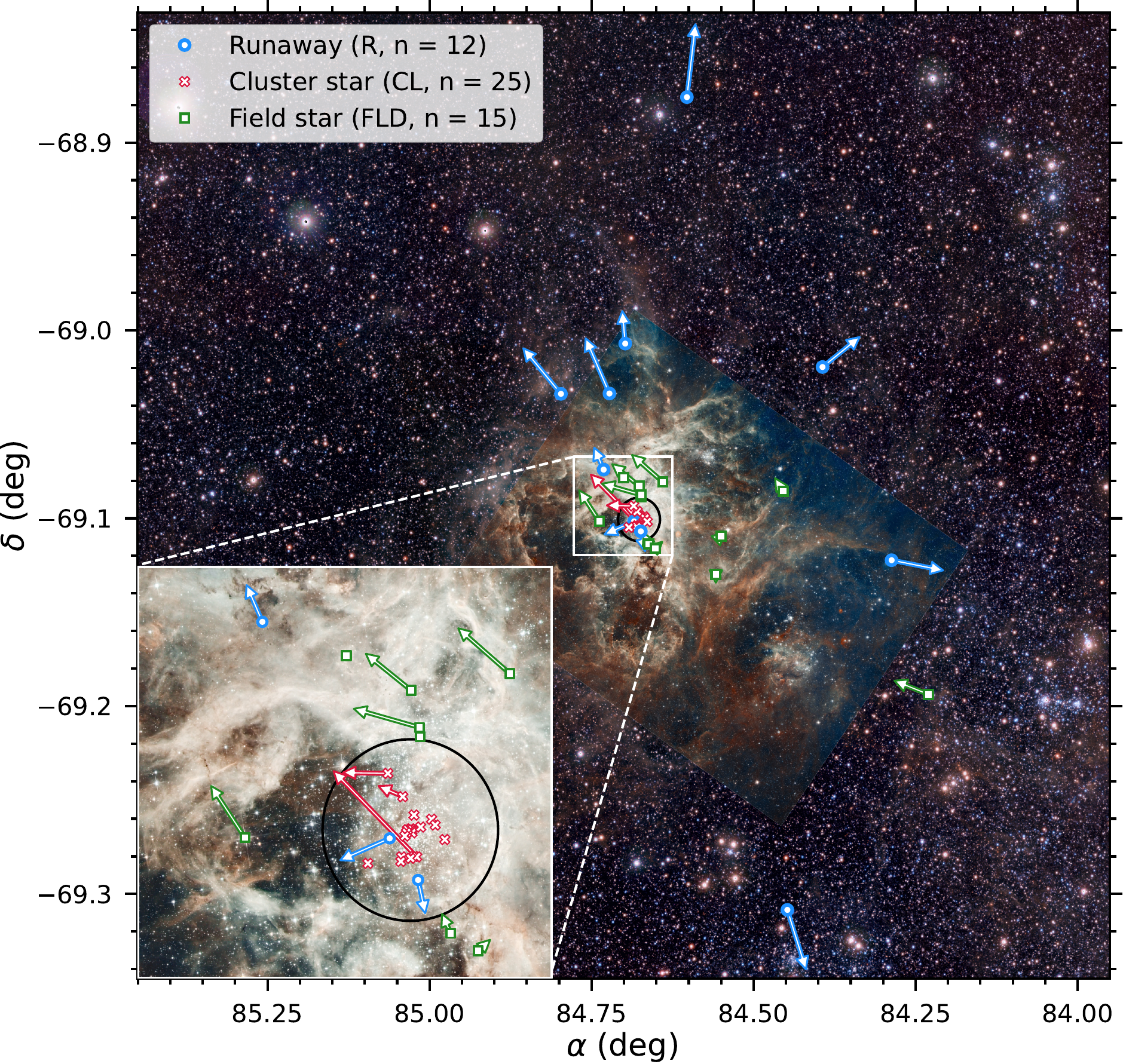}
\caption{\textbf{$\vert$ On-sky distribution of the most luminous stars (log[$L$/L$_{\odot}$] $>$ 6.0) in R136 and 30 Doradus}. Arrows depict the transverse motion direction and the length of the arrows are proportional to the transverse velocity with respect to R136. Blue circles depict runaways coming from R136, red crosses denote stars which are classified as member of R136 ($r_{\rm{proj}}$ $<$ 10 pc), and green squares indicate stars which are neither runaways originating from R136 nor member of R136. The foreground image is a composite image from the \textit{Hubble Space Telescope} (\textit{HST}) and European Southern Observatory (ESO) 2.2m telescope (NASA, ESA \& Lennon et al.; 2012). The background shows a near-infrared image of the Tarantula Nebula produced by the ESO Visible and Infrared Survey Telescope for Astronomy (VISTA, credit: ESO/M.-R. Cioni/VISTA Magellanic Cloud survey).}
\label{fig:dor30_log6_luminous}
\end{figure*}

\renewcommand{\tablename}{Extended Data Table}
\begin{table}
\renewcommand\thetable{1}
\centering
\caption{\textbf{$\vert$ Astrometric, kinematic, and physical parameters of R136}.}
\label{tab:R136params}
\begin{tabular}{l l l}
\toprule
\noalign{\smallskip}\multicolumn{3}{c}{Equatorial} \\
\noalign{\smallskip}\midrule
\noalign{\smallskip}Right Ascension$^{*}$ & $\alpha$ & 84.67664 deg \\
\noalign{\smallskip}Declination$^{*}$ & $\delta$ & -69.10084 deg \\
\noalign{\smallskip}Proper motion$^{\dagger}$ & $\mu_{\alpha^{*}}$ & 1.654 mas yr$^{-1}$ \\
\noalign{\smallskip}Proper motion$^{\dagger}$ & $\mu_{\delta}$ & 0.573 mas yr$^{-1}$ \\
\noalign{\smallskip}$\mu_{\alpha^{*}}$ dispersion & $\sigma_{\alpha}$ & - \\
\noalign{\smallskip}$\mu_{\delta}$ dispersion & $\sigma_{\delta}$ & - \\
\noalign{\smallskip}\midrule
\noalign{\smallskip}\multicolumn{3}{c}{Galactic} \\
\noalign{\smallskip}\midrule
\noalign{\smallskip}Galactic longitude$^{*}$ & $l$ & 279.46505 deg \\ 
\noalign{\smallskip}Galactic latitude$^{*}$ & $b$ & -31.67190 deg \\
\noalign{\smallskip}Proper motion$^{\dagger}$ & $\mu_{l^{*}}$ & -0.757  mas yr$^{-1}$ \\
\noalign{\smallskip}Proper motion$^{\dagger}$ & $\mu_{b}$ & 1.578  mas yr$^{-1}$ \\
\noalign{\smallskip}$\mu_{l^{*}}$ dispersion & $\sigma_{l}$ & - \\
\noalign{\smallskip}$\mu_{b}$ dispersion & $\sigma_{b}$ & - \\
\noalign{\smallskip}\midrule
\noalign{\smallskip}\multicolumn{3}{c}{Radial} \\
\noalign{\smallskip}\midrule
\noalign{\smallskip}Parallax \cite{Pietrzynski2019} & $\varpi$ & 0.0202 mas \\
\noalign{\smallskip}Distance \cite{Pietrzynski2019} & $d$ & 49.59 $\pm$ 0.09 ($\pm$ 0.54) kpc \\
\noalign{\smallskip}Radial velocity \cite{Sana2022} & $v_{\rm{R}}$ & 268.2 $\pm$ 8.6 km s$^{-1}$\\
\noalign{\smallskip}$v_{\rm{R}}$ dispersion \cite{HenaultBrunet2012} & $\sigma_{\rm{R}}$ & 4-5 km s$^{-1}$ \\
\noalign{\smallskip}\midrule
\noalign{\smallskip}\multicolumn{3}{c}{Physical properties} \\
\noalign{\smallskip}\midrule
\noalign{\smallskip}Mass \cite{Crowther2016} & $M_{\rm{cl}}$ & $\sim$ 5 $\cdot$ $10^{4}$ M$_{\odot}$\\
\noalign{\smallskip}Radius \cite{PortegiesZwart2010} & $r_{\rm{cl}}$ & $\sim$ 0.1 pc \\
\noalign{\smallskip}Age (runaways) & - & 1.83$^{+0.15}_{-0.10}$ Myr \\
\noalign{\smallskip}Age (literature) \cite{Brands2022} & - & 1.0-2.5 Myr \\
\noalign{\smallskip}Visual extinction \cite{Brands2023} & $A_{\rm{V}}$ & 1.70 $\pm$ 0.45 mag \\
\noalign{\smallskip}Number of O stars$^{\ddagger}$ & - & $\sim$ 4 $\cdot$ 10$^{2}$ \\
\noalign{\smallskip}\botrule
\noalign{\smallskip}
\end{tabular}
\footnotetext{$^{*}$R136a1 is assumed to be the centre of R136.}
\footnotetext{$^{\dagger}$Estimated using the runaways, see main text.}
\footnotetext{$^{\ddagger}$Extrapolating the IMF slope ($-1.95$) between 18-300 M$_{\odot}$ assuming M$_{\rm{cl}}$ = 5 $\times$ $10^{4}$ M$_{\odot}$.}
\end{table}

\renewcommand{\tablename}{Extended Data Table}
\begin{table}[h]
\renewcommand\thetable{2}
\caption{\textbf{$\vert$ All stars in 30 Doradus with log($L$/L$_{\odot}$) $>$ 6.0}. The uncertainty on the data is expressed as a 1$\sigma$ confidence interval.}
\label{tab:luminous_stars}
\tiny
\begin{tabular}{l l l l l l l}
\toprule%
\noalign{\smallskip}Identifier & Spectral type & $r_{\rm{sep,proj}}$ & log($L$/L$_{\odot}$) & v$_{\rm{R}}$ & log($Q_{0}$) & Ref. \\
\noalign{\smallskip}- & - & pc & - & km s$^{-1}$ & s$^{-1}$ & - \\
\midrule%
\noalign{\smallskip}R136a1 & WN5h & 0.0$^{*}$ & 6.86 & - & 50.71 & \cite{Crowther2016,Bestenlehner2020,Brands2022} \\
\noalign{\smallskip}R136a2 & WN5h & 0.02 & 6.71 & - & 50.59 & \cite{Crowther2016,Bestenlehner2020,Brands2022} \\
\noalign{\smallskip}R136a5 & O2 I(n)f* & 0.07 & 6.32 & - & 50.13 & \cite{Crowther2016,Bestenlehner2020,Brands2022} \\
\noalign{\smallskip}R136a7 & O3 III(f*) & 0.09 & 6.36 & - & 50.21 & \cite{Crowther2016,Bestenlehner2020,Brands2022} \\
\noalign{\smallskip}R136a3 & WN5h & 0.12 & 6.7 & - & 50.56 & \cite{Crowther2016,Bestenlehner2020,Brands2022} \\
\noalign{\smallskip}R136a8 & - & 0.12 & 6.17 & - & 49.98 & \cite{Crowther2016,Bestenlehner2020,Brands2022} \\
\noalign{\smallskip}R136a4 & O3 V((f*))(n) & 0.14 & 6.28 & - & 50.10 & \cite{Crowther2016,Bestenlehner2020,Brands2022} \\
\noalign{\smallskip}R136a6 & O2 I(n)f*p & 0.19 & 6.24 & - & 50.10 & \cite{Crowther2016,Bestenlehner2020,Brands2022} \\
\noalign{\smallskip}H36 & O2 If* & 0.38 & 6.27 & - & 50.10 & \cite{Crowther2016,Bestenlehner2020,Brands2022} \\
\noalign{\smallskip}H46 & O2-3 III(f*) & 0.43 & 6.1 & - & 49.96 & \cite{Crowther2016,Bestenlehner2020,Brands2022} \\
\noalign{\smallskip}R136b & O4 If & 0.52 & 6.35 & - & 50.00 & \cite{Crowther2016,Bestenlehner2020,Brands2022} \\
\noalign{\smallskip}R136c & WN5h & 0.90 & 6.58 & - & 50.32 & \cite{Bestenlehner2014} \\
\noalign{\smallskip}VFTS 1014 & O3 V + mid/late O & 1.22 & 6.22 & - & 50.00 & \cite{Bestenlehner2014} \\
\noalign{\smallskip}Melnick 42 & O2 If* & 1.70 & 6.56 & - & 50.37 & \cite{Bestenlehner2014} \\
\noalign{\smallskip}\textbf{Melnick 34 A} & WN5h & 2.52 & 6.43 & 287 $\pm$ 5 & 50.39$^{\dagger}$ & \cite{Doran2013,Tehrani2019} \\
\textbf{Melnick 34 B} & WN5h & 2.52 & 6.43 & 287 $\pm$ 5 & 50.39$^{\dagger}$ & \cite{Doran2013,Tehrani2019} \\
\noalign{\smallskip}VFTS 1001 & WN6h & 2.70 & 6.20 & - & 49.96 & \cite{Schnurr2008,Bestenlehner2014,Shenar2019} \\
\noalign{\smallskip}VFTS 482 AB & O2.5 If*/WN6 & 2.89 & 6.40 & 226 $\pm$ 13 & 50.14 & \cite{Bestenlehner2014,Castro2018} \\
\noalign{\smallskip}VFTS 1021 & O4 If+ & 2.96 & 6.34 & - & 50.02 & \cite{Bestenlehner2014} \\
\noalign{\smallskip}VFTS 1017 & O2 If*/WN5 & 3.04 & 6.21 & - & 50.05 & \cite{Bestenlehner2014} \\
\noalign{\smallskip}VFTS 1022 & O3.5 If*/WN7 & 3.11 & 6.48 & - & 50.22 & \cite{Bestenlehner2014} \\
\noalign{\smallskip}VFTS 545 & O2 If*/WN5 & 3.12 & 6.30 & - & 50.11 & \cite{Bestenlehner2014} \\
\noalign{\smallskip}VFTS 1028 & O3 III(f*) or O4-5 V & 3.62 & 6.09 & 280 $\pm$ 6 & 49.91 & \cite{Schnurr2008,Bestenlehner2014,Castro2018} \\
\noalign{\smallskip}VFTS 542 & O2 If*/WN5 + B0 V & 3.78 & 6.16 & 269 $\pm$ 8 & 49.94 & \cite{Schnurr2008,Bestenlehner2014} \\
\noalign{\smallskip}VFTS 468 & O2 V((f*)) + OB & 4.06 & 6.00 & - & 49.79 & \cite{Bestenlehner2014} \\
\noalign{\smallskip}\textbf{VFTS 512 AB} & O2 V-III((f*)) + ? & 5.57 & 6.04 & - & 49.86 & \cite{Sana2013,Bestenlehner2014} \\
\noalign{\smallskip}VFTS 599 & O3 III(f*) & 6.06 & 6.01 & 265.0 $\pm$ 1.3 & 49.77 & \cite{Sana2013,Bestenlehner2014,Schneider2018} \\
\noalign{\smallskip}VFTS 562 & O4 V & 6.74 & 6.05 & 278 $\pm$ 8 & 49.79 & \cite{Sana2013,Bestenlehner2014} \\
\noalign{\smallskip}VFTS 506 AB & ON2 V((n))((f*)) + ? & 10.33 & 6.24 & - & 50.05 & \cite{Sana2013,Bestenlehner2014} \\
\noalign{\smallskip}VFTS 509 AB & WN5(h) + O4 V & 11.35 & 6.09 & 220 $\pm$ 10 & 50.17 & \cite{Bestenlehner2014,Shenar2019} \\
\noalign{\smallskip}VFTS 457 & O3.5 If*/WN7 & 12.27 & 6.20 & - & 49.89 & \cite{Bestenlehner2014} \\
\noalign{\smallskip}VFTS 427 & WN8(h) & 15.38 & 6.13 & - & 49.90 & \cite{Bestenlehner2014} \\
\noalign{\smallskip}VFTS 527 A & O6.5 I & 15.38 & 6.20 & 262.4 $\pm$ 0.1 & 49.78$^{\dagger}$ & \cite{Doran2013,Mahy2020} \\
\noalign{\smallskip}VFTS 527 B & O7 I & 15.38 & 6.20 & 262.4 $\pm$ 0.1 & 49.78$^{\dagger}$ & \cite{Doran2013,Mahy2020} \\
\noalign{\smallskip}VFTS 695 A & WN6h & 18.85 & 6.35 & 270 $\pm$ 5 & 49.74$^{\dagger}$ & \cite{Doran2013,Shenar2019} \\
\noalign{\smallskip}VFTS 695 B & O3.5 If/WN7 & 18.85 & 6.35 & 270 $\pm$ 5 & 49.74$^{\dagger}$ & \cite{Doran2013,Shenar2019} \\
\noalign{\smallskip}VFTS 621 & O2 V((f*))z & 20.54 & 6.14 & - & 49.97 & \cite{Bestenlehner2014} \\
\noalign{\smallskip}VFTS 402 & WN5(h) + WN7(h): & 20.62 & 6.07 & 274 $\pm$ 9 & 49.69 & \cite{Bestenlehner2014,Shenar2019} \\
\noalign{\smallskip}\textbf{VFTS 682} & WN5h & 28.45 & 6.51 & 300 $\pm$ 10 & 50.35 & \cite{Bestenlehner2014} \\
\noalign{\smallskip}VFTS 259 A & O6 Iaf & 39.82 & 6.1 & - & 49.71 & \cite{Bestenlehner2014} \\
\noalign{\smallskip}VFTS 267 A & O3 III-I(n)f* & 44.5 & 6.01 & - & 49.79 & \cite{Bestenlehner2014} \\
\noalign{\smallskip}\textbf{R144 A} & WN5/6h & 59.67 & 6.44 & 210 $\pm$ 20 & 49.95$^{\dagger}$ & \cite{Doran2013,Shenar2021} \\
\textbf{R144 B} & WN6/7h & 59.67 & 6.44 & 210 $\pm$ 20 & 49.95$^{\dagger}$ & \cite{Doran2013,Shenar2021} \\
\noalign{\smallskip}\textbf{VFTS 758} & WN5h & 68.75 & 6.36 & - & 50.17 & \cite{Bestenlehner2014} \\
\noalign{\smallskip}R130 A & WC4 & 69.88 & 6.01 & 332 $\pm$ 7 & 49.60$^{\dagger}$ & \cite{Doran2013,Shenar2019} \\
R130 B & B1Ia & 69.88 & 6.07 & 332 $\pm$ 7 & 49.60$^{\dagger}$ & \cite{Doran2013,Shenar2019} \\
\noalign{\smallskip}\textbf{VFTS 617} & WN5ha & 83.36 & 6.29 & - & 50.14 & \cite{Bestenlehner2014} \\
\noalign{\smallskip}\textbf{VFTS 72} & O2 V-III(n)((f*)) & 112.21 & 5.96-6.06 & 273.6 $\pm$ 2.2 & 49.75-49.93 & \cite{Bestenlehner2014,SabinSanjulian2017} \\
\noalign{\smallskip}\textbf{VFTS 16} & O2 III-If* & 121.67 & 6.12 & 189.4 $\pm$ 1.3 & 50.08 & \cite{Bestenlehner2014} \\
\noalign{\smallskip}VFTS 3 & B1 Ia+ & 159.48 & 6.03 & - & 47.62$^{\ddagger}$ & \cite{McEvoy2015} \\
\noalign{\smallskip}\textbf{MCPS 084.44781-69.30846} & O2-3 V-III((f*)) & 193.12 & 5.5-6.1$^{\S}$ & 277.9 $\pm$ 2.9 & 49.3-49.9$^{\S}$ & \cite{Evans2015b} \\
\noalign{\smallskip}\textbf{SK-68 137} & O2-3.5 III(f*) & 195.85 & 6.19 & 273 & 50.07$^{\ddagger}$ & \cite{Walborn1995,Walborn2004} \\
\botrule%
\end{tabular}
\footnotetext{$^{*}$R136a1 is assumed to be the centre of R136.}
\footnotetext{$^{\dagger}$Each component in the binary is assumed to contribute half of the total ionising photons.}
\footnotetext{$^{\ddagger}$Estimated in this work, see main text.}
\footnotetext{$^{\S}$Range is given by the lower and upper values of stars with similar spectral types in R136.}
\end{table}

\clearpage

\section*{Supplementary Information}

\subsection*{Selection biases}
\label{sec:selection_biases}
We have searched for massive runaways coming from R136. In the process of doing so, we have introduced filters on the accuracy of the parallax and on the velocity of the runaways. These could lead to biases in the observed runaway sample, for example making it so we preferentially find brighter and thus more massive stars to be runaways. In turn, this could lead to a shallower initial mass function.

To investigate this further, we show the parallax uncertainty as a function of the G-magnitude in Sup. Fig.~\ref{fig:bias_parallax} for the 55 observed runaways (Sup. Table~\ref{tab:kinematic_runaways}), the 9,368 blue candidate runaways (Bp - Rp $<$ 1.0 mag), and the 24,793 blue stars possibly located at the distance of the LMC (Bp - Rp $<$ 1.0 mag, $\varpi$ $<$ 0.15 mas, \texttt{ruwe} $<$ 1.4). We exclude stars with $\sigma_{\varpi}$ $<$ 0.05 mas. This cut-off leads us to reach $\sim$ 95\% completeness between G $\sim$ 16.2-16.4 mag and $\sim$ 50\% completeness between G $\sim$ 16.8-17.0 mag. The former corresponds to a 16 M$_{\odot}$ and the latter to a 12 M$_{\odot}$ main sequence star at the distance of the LMC assuming $A_{\rm{V}}$ = 1.2 mag. For fainter magnitudes, a significant fraction of stars are excluded because of this cut-off. We can therefore miss out on late-type O stars with relatively large $A_{\rm{V}}$ and a significant fraction of the early B-type runaways if present. Correcting for this bias would require us to know the intrinsic luminosity function and therefore the mass function of the runaways, which is exactly what we are trying to determine. While we could loosen the restriction on the parallax uncertainty, this would add additional biases as we would start to include a significant population of foreground stars due to their more uncertain distance. We therefore restrict the determination of the mass function of the runaways to stars with a mass larger than 16 M$_{\odot}$.

We show the transverse velocity as a function of the uncertainty on the transverse velocity in Sup. Fig.~\ref{fig:bias_velocity} for the 55 observed runaways, 9,368 blue candidate runaways (Bp - Rp $<$ 1.0 mag), and the 24,793 blue stars possibly at the distance of the LMC (Bp - Rp $<$ 1.0 mag, $\varpi$ $<$ 0.15 mas, \texttt{ruwe} $<$ 1.4). We exclude stars with $v_{\rm{T}}$ / $\sigma_{v_{\rm{T}}}$ $<$ 3 indicated with the red region. The candidate runaways occupy this region as well, indicating that we could miss out on runaways which have a large transverse velocity uncertainty. The runaway candidates which we are now excluding have a transverse velocity between 27.6 and $\sim$ 55 km s$^{-1}$ with respect to R136. Out of the 9,368 candidate runaways, 4,108 have $v_{\rm{T}}$ $<$ 55 km s$^{-1}$. Out of these 4,108 candidate runaways, 3,097 and 1,011 have $v_{\rm{T}}$ / $\sigma_{v_{\rm{T}}}$ larger and smaller than 3, respectively.

We find 23 runaways with a transverse velocity between 27.6 and 55 km s$^{-1}$ among these 3,097 early-type candidate runaways with $v_{\rm{T}}$ / $\sigma_{v_{\rm{T}}}$ $>$ 3 and $v_{\rm{T}}$ $<$ 55 km s$^{-1}$. If the ratio of observed runaways to runaway candidates is similar, we can estimate the number of runaways missed out on. We observe 23 runaways with a transverse velocity smaller than 55 km s$^{-1}$ among the 3,097 early-type candidate runaways. The 1,011 early-type candidate runaways now excluded are estimated to contain $\sim$ 8 true runaways. Out of the 1,011 excluded runaway candidates with $v_{\rm{T}}$ / $\sigma_{v_{\rm{T}}}$ $<$ 3, 95\% have G between 15.7-17.3 mag, implying that we can miss out on late-type O stars and early B-type stars. We apply a correction to the initial mass function of the runaways below. The initial mass function of R136 remains unchanged as these have a mass larger than 30 M$_{\odot}$.

\renewcommand{\figurename}{Sup. Fig.}
\renewcommand\thefigure{1}
\begin{figure}[h]
\centering
\includegraphics[width=0.99\linewidth]{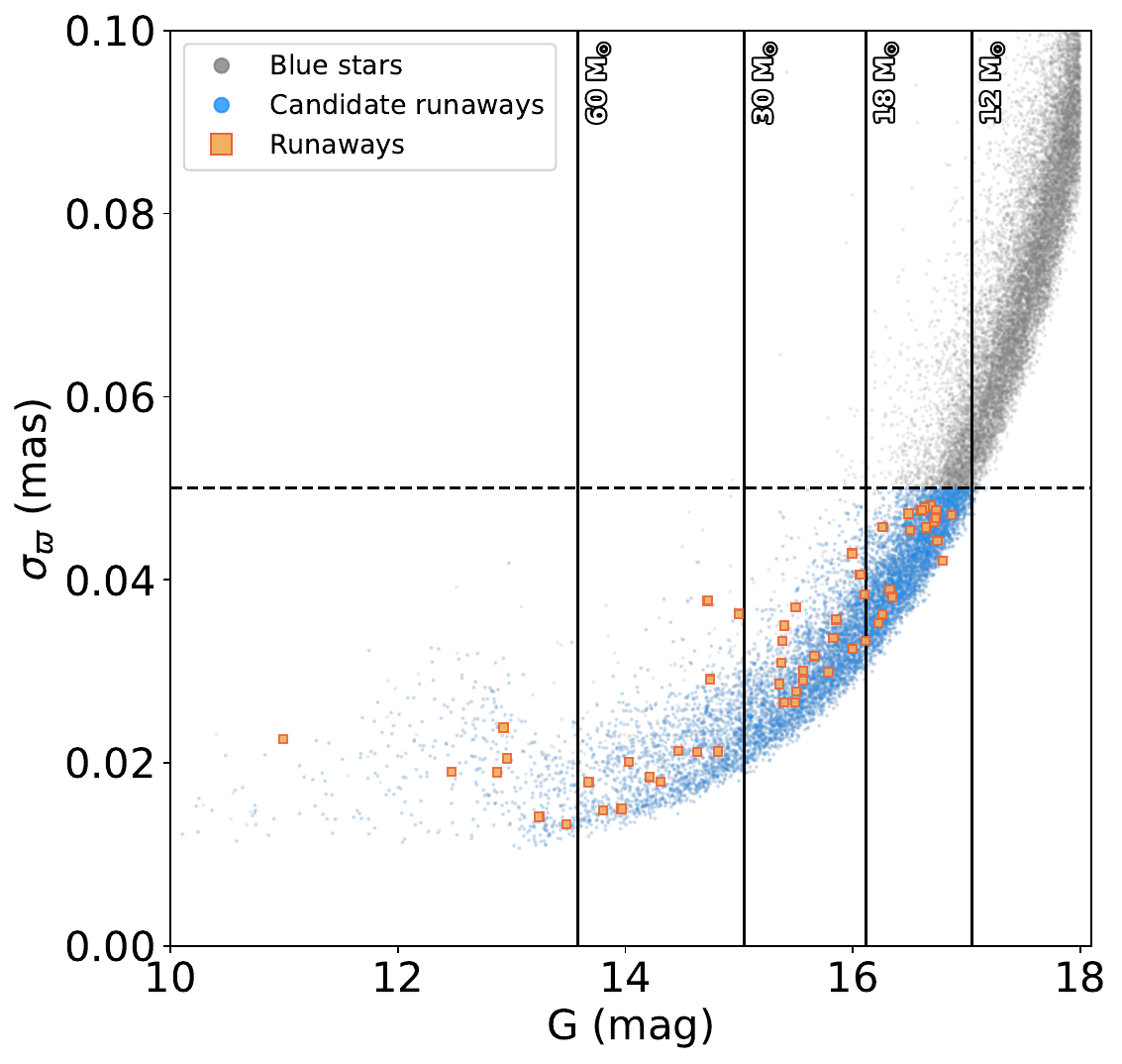}
\caption{\textbf{$\vert$ Parallax uncertainty as a function of the G-magnitude}. The 55 runaways are shown with the yellow squares, the 9,368 blue runaway candidates as the blue dots, and the 24,793 blue stars located beyond $\sim$ 6.7 kpc with the grey dots (Bp - Rp $<$ 1.0 mag, $\varpi$ $<$ 0.15 mas, and \texttt{ruwe} $<$ 1.4). The black dashed horizontal line denotes the parallax uncertainty cut-off (0.05 mas) adopted in this work. The four vertical black solid lines indicate the expected G-magnitude of stars with a mass of 60, 30, 18, 10 M$_{\odot}$ at the distance of the LMC and an assumed $A_{\rm{V}}$ = 1.2 mag.}
\label{fig:bias_parallax}
\end{figure}

\renewcommand{\figurename}{Sup. Fig.}
\renewcommand\thefigure{2}
\begin{figure}[h]
\centering
\includegraphics[width=0.99\linewidth]{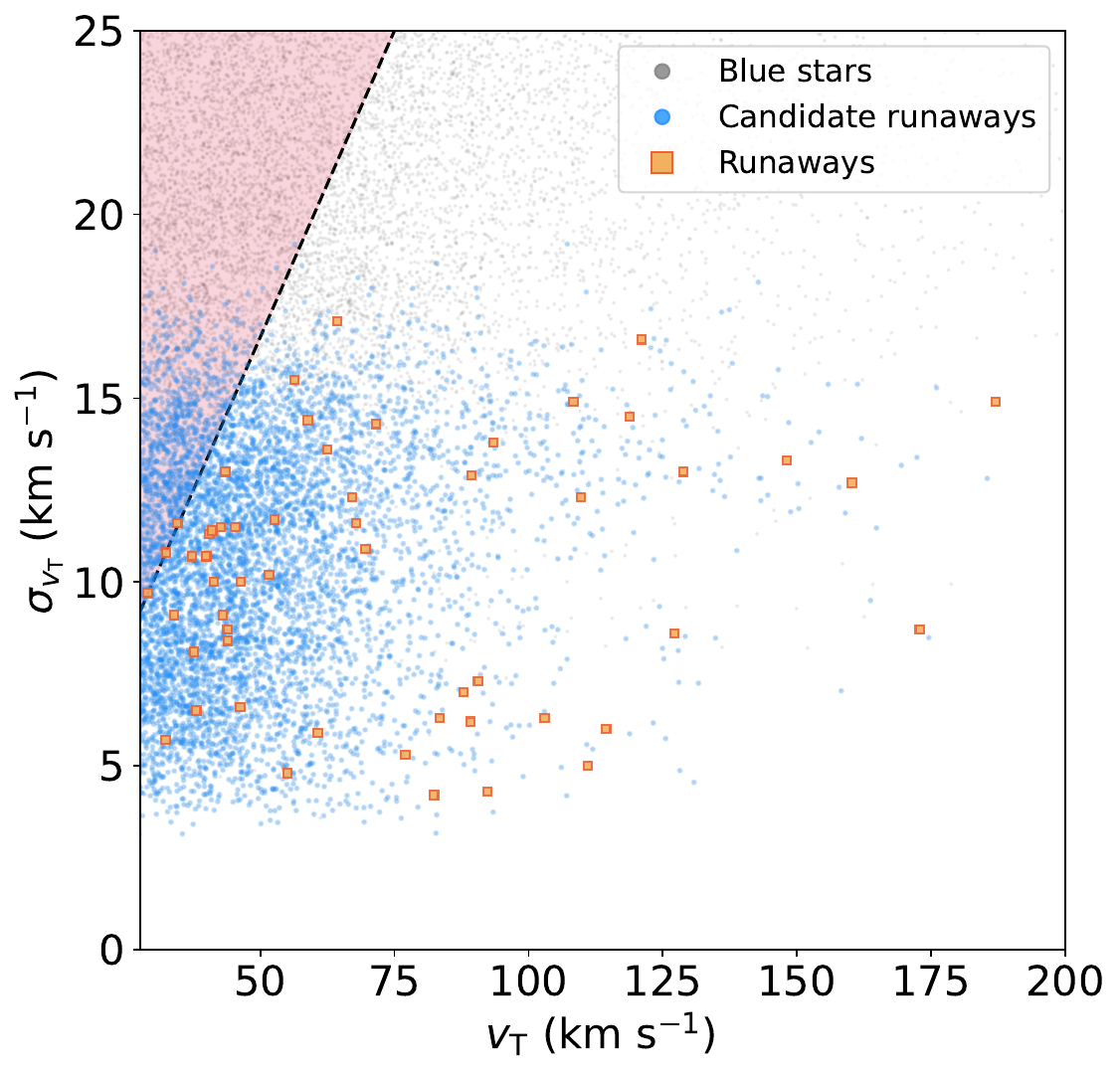}
\caption{\textbf{$\vert$ Transverse velocity uncertainty as a function of the transverse velocity}. The 55 runaways are shown with the yellow squares, the 9,368 blue runaway candidates as the blue dots, and the 24,793 blue stars located beyond $\sim$ 6.7 kpc with the grey dots (Bp - Rp $<$ 1.0 mag, $\varpi$ $<$ 0.15 mas, and \texttt{ruwe} $<$ 1.4). The black dashed diagonal line indicates the threshold where $\sigma_{v_{\rm{T}}}$ / $v_{\rm{T}}$ = 3. The red region above this line denotes where we exclude runaways due to their uncertain transverse velocity.}
\label{fig:bias_velocity}
\end{figure}

\clearpage

\subsection*{Initial mass function of R136}
\label{sec:imf_r136}
We fit the initial mass function (IMF) of the R136 cluster twice: first by taking into account only the massive stars in the core of the cluster \citep[as analysed by][]{Brands2022}, and second by taking into account both the stars in the cluster core and the massive runaways associated with the cluster, that we identified in the current work.

The initial mass of the runaways is taken from the literature (see Sup. Table~\ref{tab:luminous_stars}). For stars with a current mass below 30 M$_{\odot}$ we assume that the current mass is the initial mass, as a negligible amount of mass has been lost by their stellar wind within a few million years. For MCPS 084.44781-69.30846, SK-68 137, and VFTS 512 we estimated the initial mass by taking the average of the initial mass of the stars with similar spectral type in R136 \citep{Brands2022}. For fitting the IMF slope we first build a probability distribution for the cluster. We do this by adding the probability functions of the initial mass of the individual stars (where in one case we do not include the runaways, and in the other case we do). We assume a Gaussian probability distribution for each star, where the centre of the distribution lies at the most likely value of  the initial mass and the width is determined by the $1\sigma$ uncertainty. 
Once we have constructed the probability function, we obtain the slope of the IMF by normalising the probability function to unity and fitting a linear function to it in log-log space, considering only the part between 30~$M_{\odot}$ and 300~$M_{\odot}$. 

We use only the primary masses of the stars and runaways considered. Secondary masses are only known for Melnick 34 \citep[$M_{\rm{sec}}$ = 127 $\pm$ 17 M$_{\odot}$;][]{Tehrani2019} and RMC 144 \citep[$M_{\rm{sec}}$ = 100 $\pm$ 12 M$_{\odot}$;][]{Shenar2021} which would produce an even shallower slope than determined below.

When only considering the stars that are currently residing in the cluster core, we find a best fitting IMF slope of $\gamma_{\rm{R136}} = -2.32\pm0.16$; when we add the runaways we find $\gamma_{\rm{R136}} = -1.95\pm0.08$. 
The probability functions as well as the best fits are shown in Sup. Fig.~\ref{fig:r136_IMF}. In this figure we show also the uncertainties on the probability function, that we obtained by bootstrapping. These uncertainties are not considered in the fitting process. This is because the uncertainties are larger in the high-mass range compared to the lower-mass range: in a weighted fit the high-mass stars ($\gtrsim 100~M_{\odot}$) would carry very little importance and the resulting value of $\gamma_{\rm{R136}}$ would reflect the IMF slope at lower masses only.

We performed a similar test excluding the runaways with $t_{\rm{evo}} > 3.0$~Myr, which could originate from the sub-cluster encounter. These stars can be considered to be born in the sub-cluster and not in R136 and should therefore not be included in the IMF of R136. The combination of the stars in R136 with the runaways with $t_{\rm{evo}}$ $<$ 3.0 Myr gives an even shallower $\gamma_{\rm{R136}} = -1.83\pm0.09$. On the other hand, we can also consider all runaways with kinematic age $t_{\rm{kin}} < 1.0$~Myr to be part of the sub-cluster encounter and exclude them from the IMF determination. The combination of the stars in R136 with the runaways with $t_{\rm{kin}}$ $>$ 1.0 Myr gives $\gamma_{\rm{R136}} = -2.27\pm0.16$.

We fit the initial mass function of only the runaways with a similar method. As discussed in Section~\ref{sec:selection_biases}, we likely missed out on about eight runaways. These runaways are expected to have masses between 10-20 M$_{\odot}$ and would make the intrinsic mass function of the runaways steeper than the observed one. To correct for this, we randomly select eight stars from the 2128 candidate runaways with $\sigma_{v_{\rm{T}}}$ / $v_{\rm{T}}$ $<$ 3 (see Sup. Fig.~\ref{fig:bias_velocity}) and add them to the observed runaway sample. The initial mass function of the runaways with only determined masses is $\gamma_{\rm{rw,M}} = -0.93\pm0.09$. With optical photometry we estimate the mass of the runaways for which we do not have spectral information. We calculate the absolute G-magnitude at the distance of the LMC assuming A$_{\rm{V}}$ = 1.2 mag, the average visual extinction towards the observed runaways. We approximate M$_{\rm{G}}$ $\sim$ M$_{\rm{V}}$ and find the mass of the OB-type dwarf with the nearest M$_{\rm{V}}$ in \cite{Martins2005} and \cite{Pecaut2013}. For MCPS 084.44781-69.30846, SK-68 137, and VFTS 512 we again estimate the initial mass by taking the average of the initial mass of the stars with similar spectral type in R136 \citep{Brands2022}. We repeat this process 100 times, which results in an average slope of the mass function $\gamma_{\rm{rw,all}} = -1.87\pm0.07$ between 16 and 200 M$_{\odot}$. Thus, the IMF of the runaways is shallower than the Salpeter IMF indicating that runaways are predominantly massive stars.

\renewcommand{\figurename}{Sup. Fig.}
\renewcommand\thefigure{3}
\begin{figure}
\centering
\includegraphics[width=0.99\linewidth]{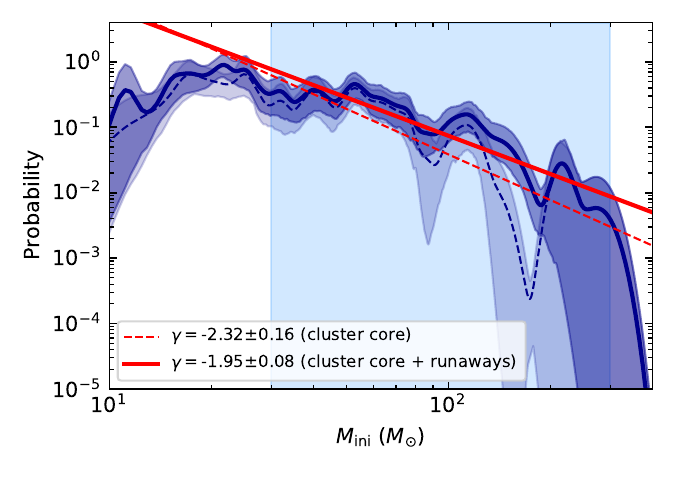}
\caption{\textbf{$\vert$ Determination of the initial mass function of R136 with and without runaways}. The dashed dark blue line shows the probability function of only the cluster core; the solid dark blue line the probability function of the cluster core together with the runaways identified in this work. Shaded regions around the probability function indicate the bootstrapped uncertainty on the distributions. In red we show the best fit power law slope of the IMF. The light blue shading in the back indicates the mass range that was considered in the fitting process. The uncertainty on the IMF is expressed as a 1$\sigma$ confidence interval.}
\label{fig:r136_IMF}
\end{figure}

\clearpage

\subsection*{Motion of runaways}
\label{sec:results_motion_runaways}

\renewcommand{\figurename}{Sup. Fig.}
\renewcommand\thefigure{4}
\begin{figure}
\centering
\includegraphics[width=0.99\linewidth]{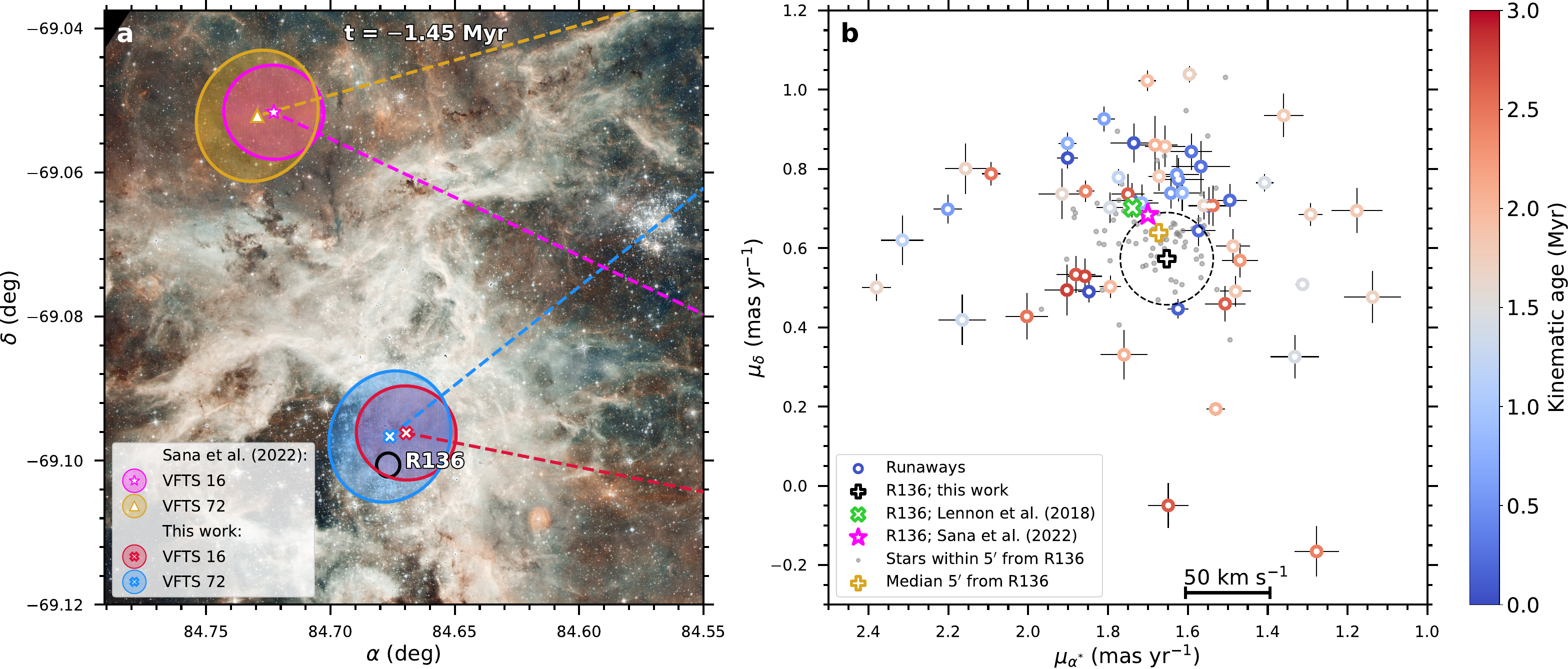}
\caption{\textbf{$\vert$ The proper motion and trace-back of runaways coming from R136.} \textbf{a}: Trace-back of VFTS 16 and 72 for $t$ = --1.45 Myr for the R136 proper motion determined in this work (blue and red, respectively) and for the proper motion determined in \citet{Sana2022} (magenta and gold, respectively). The shaded regions indicate the 1$\sigma$ uncertainty on the position for $t$ = --1.45 Myr, with the dashed lines indicating their travel path relative to R136. The background composite optical and near-infrared image is taken by the \textit{Hubble Space Telescope} (\textit{HST}) and European Southern Observatory (ESO) 2.2m telescope (NASA, ESA \& Lennon et al.; 2012). \textbf{b}: Proper motion distribution of the 55 runaways coming from R136 in the last 3 Myr. The runaways are shown with open circles coloured according to their trace-back kinematic age. Our determined R136 proper motion is shown with the black plus, the literature R136 proper motion with the green cross \citep{Lennon2018} and magenta star \citep{Sana2022}. Stars within 5$^{\prime}$ with reliable astrometry are depicted with grey dots, the median is shown with the golden plus. The uncertainty on the data is expressed as a 1$\sigma$ confidence interval.}
\label{fig:r136_runaways_pm_vfts16_72}
\end{figure}

We show the proper motion of the 55 found runaways relative to R136 in the left panel in Sup. Fig.~\ref{fig:r136_runaways_pm_vfts16_72} as open circles. They are coloured according to their trace-back kinematic age, where bluer colours indicate stars that are ejected more recently ($\lesssim$ 1.5 Myr) compared to redder colours depicting stars ejected longer ago ($\gtrsim$ 1.5 Myr). The proper motion of R136 determined in this work is shown with the black plus symbol. The runaway velocity threshold of 27.6 km s$^{-1}$ is shown with the dashed black circle. It becomes apparent that most of the runaways with a kinematic age less than 1.0 Myr (blue colours) are ejected towards the north and thus have more positive $\mu_{\delta}$ compared to R136. The runaways with grey and redder colours are ejected isotropically relative to R136.

Our proper motion determined for R136 is offset ($\sim$ 0.12-0.15 mas yr$^{-1}$; equivalent to $\sim$ 28-36 km s$^{-1}$) from other estimates in the literature; shown with the green cross for \cite{Lennon2018} and the magenta star for \cite{Sana2022}. We show the stars, within 5$^{\prime}$ from the centre of R136 (with reliable astrometry following the filters and corrections in Section~\ref{sec:methods_gaia_filters}) with the grey dots and the median of this with the golden plus symbol. This median is in between the aforementioned literature values and our determined proper motion. 

To illustrate the difference between our newly determined proper motion of R136 and that proposed by \cite{Sana2022}, we show the trace back for the two previously known runaways VFTS 16 and 72 in the right panel in Figure~\ref{fig:r136_runaways_pm_vfts16_72}. VFTS 16 and 72 are both O2-type stars with masses of 80-100 $M_{\odot}$ and were likely ejected from R136 considering their estimated evolutionary age of 0 to 1 Myr for both \citep{Gvaramadze2010,Evans2010,Lennon2018,Schneider2018}. Their error ellipses on the trace-back position overlap nearly perfectly for $t$ = --1.45 Myr, suggesting that they may have been ejected in the same dynamical interaction. We show the trace-back with respect to R136 under the assumption of the proper motion determined in this work with the blue and red ellipses and the proper motion determined in \cite{Sana2022} with the magenta and gold dashed ellipses.

We can see that for our determined R136 proper motion, the position of VFTS 16 and 72 are consistent with the position of R136 within 1$\sigma$. On the other hand, if we adopt the proper motion from \cite{Sana2022} the position of VFTS 16 and 72 deviates by 0.05 deg and $\sim$ 5$\sigma$ given their on-sky uncertainty and never crosses R136. This lends support to our determination of the R136 proper motion, which can excellently explain the ejection of VFTS 16 and 72 from the centre of R136. In conclusion, the runaways provide a way to measure the motion of R136 since the time of its formation.

\subsection*{False positive rate}
Our adopted velocity threshold of 27.6 km s$^{-1}$ is required to ensure that we do not find any members of R136 which have a velocity that extends into the tail of the Maxwell-Boltzmann distribution. The field stars around R136 have a 1D velocity dispersion larger than that of R136 \citep[5.8 km s$^{-1}$ compared to the 3.9 km s$^{-1}$ in R136;][]{Sana2022}. This could lead us to find runaways which may coincidentally have a large transverse velocity despite not being a runaway ejected from R136. We determine here the expected false positive rate for the 55 observed runaways.

To do this we resort to a Monte Carlo simulation. Out of the 21,382 stars, 9,368 have Bp - Rp $<$ 1.0 mag, indicating that they are likely early-type stars. We give these 9,368 early-type runaway candidates a random velocity in $l$ and $b$ according to the 1D velocity dispersion measured in \citep{Sana2022}. With the randomly sampled transverse velocity, we search for runaways with the same method. This is redone 10$^{4}$ times after which we find a total of 265 false-positive runaways. The expected number of false-positive runaways is therefore $\sim$ 0.03. The chance for a single star to be a false positive is 2.8 $\times$ 10$^{-6}$. With a total of 55 runaways the chance that one of these is a false positive is 1.6 $\times$ 10$^{-4}$. This does not take into account that most of our runaways have transverse velocities well above threshold of 27.6 km s$^{-1}$, while the false-positive runaways are expected to have a transverse velocity close to 27.6 km s$^{-1}$ because of the input velocity distribution.

Another source of false-positive runaways could be the statistical uncertainty on the transverse velocity. For each runaway, we determine the probability that the statistical uncertainty causes the observed transverse velocity. For each runaway we randomly sample the intrinsic transverse velocity from the field distribution of $5.8\sqrt{2}$ km s$^{-1}$ \citep{Sana2022}. We determine the probability that the observed transverse velocity is caused by the statistical uncertainty on transverse velocity. Following this, we determine the total probability that one of the runaways could be a slow-moving field star. We repeat this process 10,000 times. The average probability that one of the 55 runaways is a false positive and is instead a slow-moving field star is $p$ = $2\times10^{-2}$.

\clearpage

\subsection*{Statistical analysis}
\subsubsection*{Consistency tests}
Here we provide statistical evidence for two waves of runaways rather than a constant ejection. The objective is to investigate the consistency of the observed kinematic age distribution of the runaways with a hypothesis of constant ejection-rate, we implement the ConTEST method, as detailed in \cite{Stoppa2023}.
The ConTEST method involves the computation of a test statistic that measures the distance between the estimated probability density functions (PDFs) of two samples. In our case, these samples are the observed kinematic age distribution of the runaways and a theoretical distribution assuming a constant ejection-rate.

To evaluate the hypothesis of consistency between the observed data and the theoretical model, ConTEST creates a series of simulated samples drawn from the model's distribution, each mirroring the size of the observed dataset. For each simulated sample, it computes the same test statistic that quantifies the distance of the simulated sample to the model density. This process is rooted in the principle that if the observed sample genuinely aligns with the model, then its calculated distance should not significantly differ from those derived from the simulated samples. Essentially, if the distance measure of the observed and simulated samples significantly differs, it indicates a lack of statistical consistency with the theoretical model, leading to the rejection of the hypothesis that the observed phenomena conform to the model's expectations. 

We find that the kinematic age distribution of the runaways can not be explained by a model assuming a constant ejection-rate with a $p$-value = 0.0115 (2.5$\sigma$) (Sup. Fig.~\ref{fig:contest_tkin}). We perform a similar analysis on the distribution of ejection angles. We test the consistency of the observed ejection angles of the runaways with a hypothesis of an isotropic ejection mechanism. We find that the observed ejection angles of the runaways with kinematic age smaller than 1.0 Myr can not be explained by a model assuming isotropic ejections with a $p$-value = 0.0007 (3$\sigma$) (Sup. Fig.~\ref{fig:contest_ejang_low}). On the other hand, the observed ejection angles of the runaways with kinematic age larger than 1.0 Myr is consistent with the model assuming isotropic ejections with a $p$-value = 0.38 (Sup. Fig.~\ref{fig:contest_ejang_high}). Regardless, if we perform this analysis on the ejection angles of all runaways we find a $p$-value = 0.001. 

Last, we perform the same analysis on the distribution of transverse velocities. We test the consistency of the observed ejection angles of the runaways with a hypothesis of a power-law (N($v_{\rm{T}}$) $\sim$ N(0)$v_{\rm{T}}^{\gamma}$) with slope $\gamma$ = -1.5, predicted by dynamical ejection models \citep{Fujii2011,Perets2012,Oh2016}. We adopt a lower bound at 27.6 km s$^{-1}$ below which we do not search for runaways, and an upper bound of 250 km s$^{-1}$. We find that the observed transverse velocity distribution is consistent with this model, with a $p$-value = 0.27 (Sup. Fig.~\ref{fig:contest_vt}).

\renewcommand{\figurename}{Sup. Fig.}
\renewcommand\thefigure{5}
\begin{figure}[h]
\centering
\includegraphics[width=0.99\linewidth]{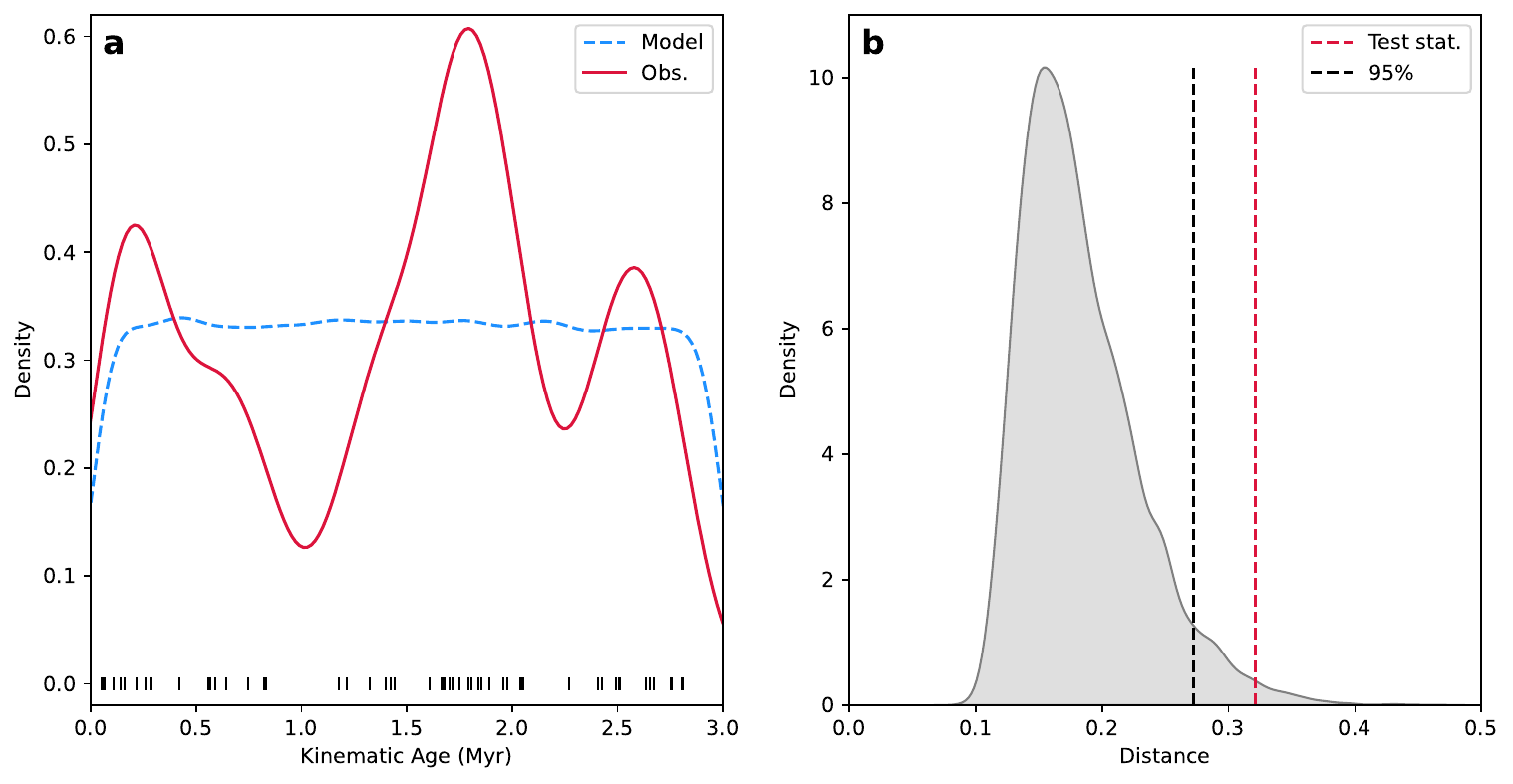}
\caption{\textbf{$\vert$ Consistency test for the kinematic age distribution of runaways}. \textbf{a}. The observed kinematic ages are shown with the black stripes. The probability density function of the observed kinematic ages is shown in red. The model assuming a constant ejection-rate is shown in blue. \textbf{b}. The distribution of the distance statistic of the simulated model sample is shown in grey. The dashed black line indicates the 95\% threshold of the sample. The dashed red line shows the distance statistic for the observed kinematic ages ($p$-value = 0.0115).}
\label{fig:contest_tkin}
\end{figure}

\renewcommand{\figurename}{Sup. Fig.}
\renewcommand\thefigure{6}
\begin{figure}[h]
\centering
\includegraphics[width=0.99\linewidth]{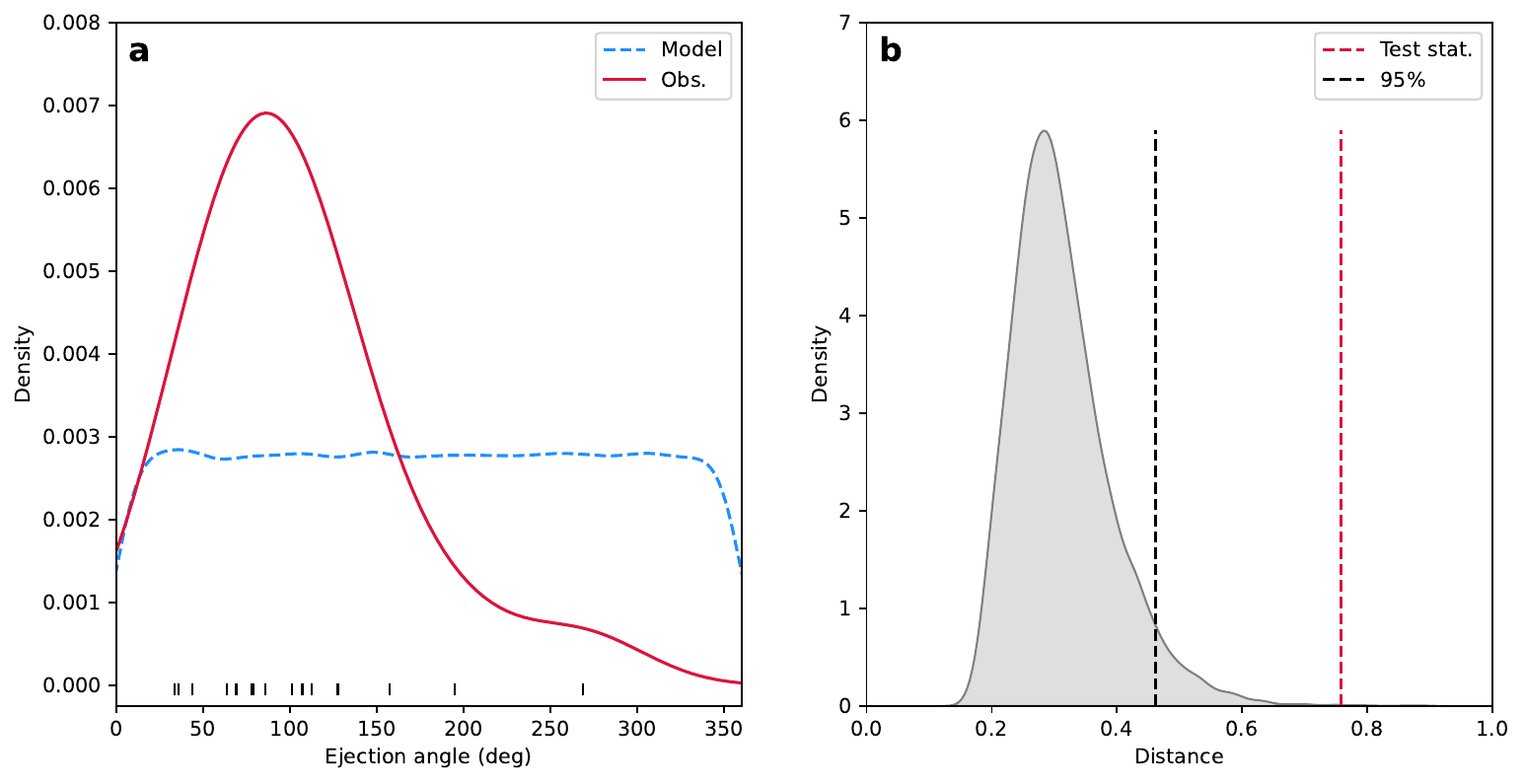}
\caption{\textbf{$\vert$ Consistency test for the ejection angle distribution of runaways with kinematic age smaller than 1.0 Myr}. \textbf{a}. The observed ejection angles are shown with the black stripes. The probability density function of the observed ejection angles is shown in red. The model assuming an isotropic ejection mechanism (e.g. uniform distribution of ejection angles) is shown in blue. \textbf{b}. The distribution of the distance statistic of the simulated model sample is shown in grey. The dashed black line indicates the 95\% threshold of the sample. The dashed red line shows the distance statistic for the observed ejection angles ($p$-value = 0.0007).}
\label{fig:contest_ejang_low}
\end{figure}

\renewcommand{\figurename}{Sup. Fig.}
\renewcommand\thefigure{7}
\begin{figure}[h]
\centering
\includegraphics[width=0.99\linewidth]{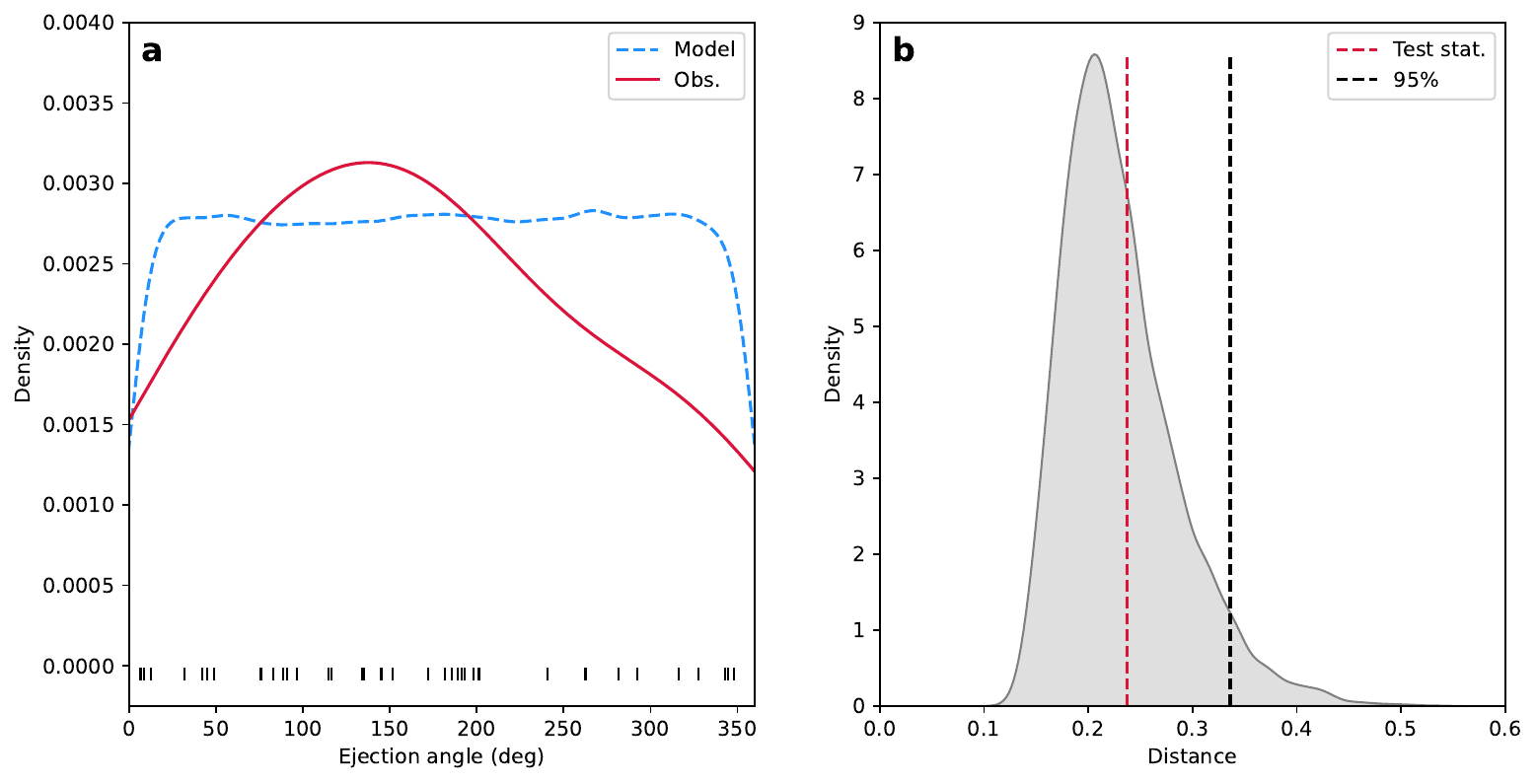}
\caption{\textbf{$\vert$ Consistency test for the ejection angle distribution of runaways with kinematic age larger than 1.0 Myr}. \textbf{a}. The observed ejection angles are shown with the black stripes. The probability density function of the observed ejection angles is shown in red. The model assuming an isotropic ejection mechanism (e.g. uniform distribution of ejection angles) is shown in blue. \textbf{b}. The distribution of the distance statistic of the simulated model sample is shown in grey. The dashed black line indicates the 95\% threshold of the sample. The dashed red line shows the distance statistic for the observed ejection angles ($p$-value = 0.38).}
\label{fig:contest_ejang_high}
\end{figure}

\renewcommand{\figurename}{Sup. Fig.}
\renewcommand\thefigure{8}
\begin{figure}[h]
\centering
\includegraphics[width=0.99\linewidth]{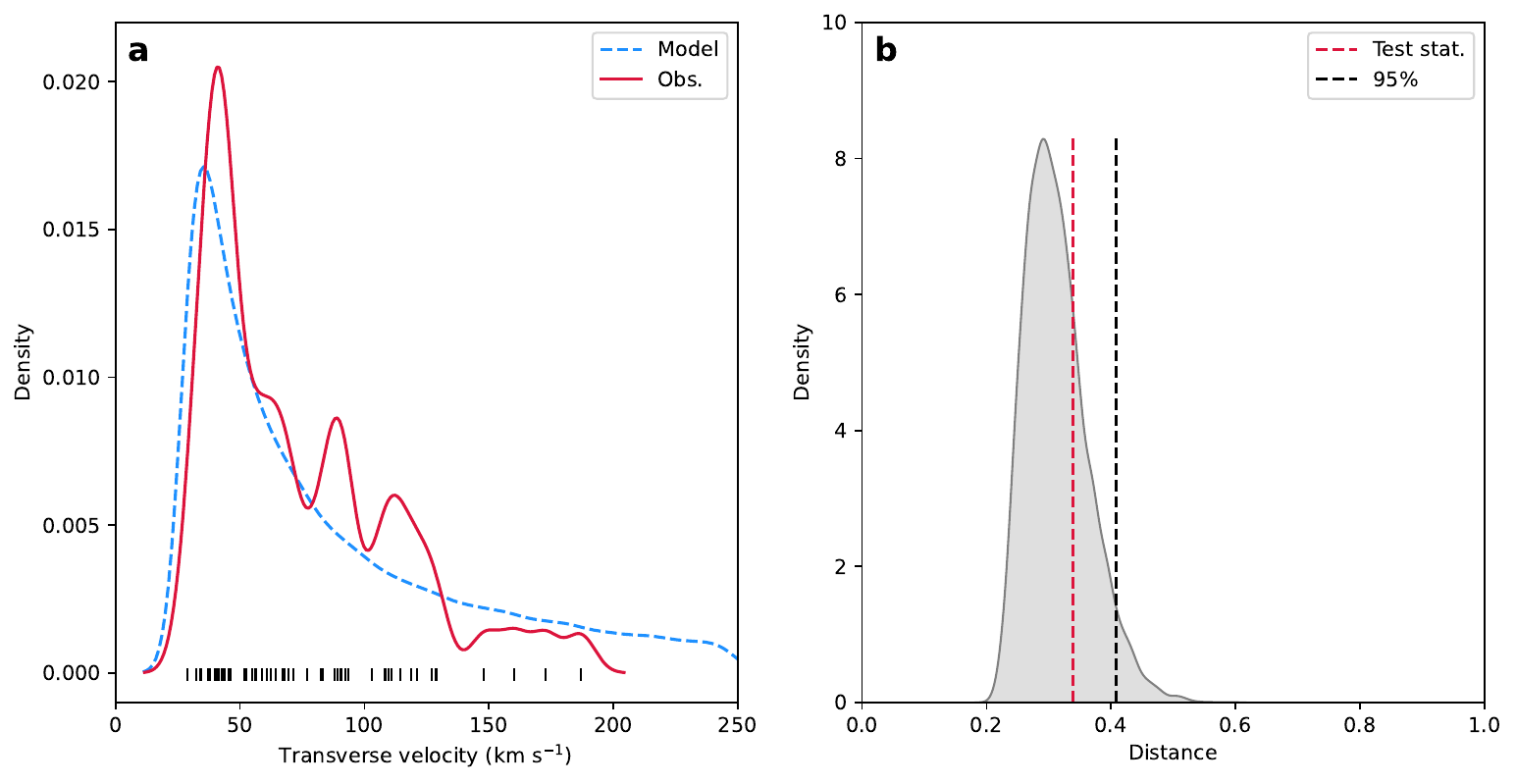}
\caption{\textbf{$\vert$ Consistency test for the transverse velocity distribution of runaways}. \textbf{a}. The observed transverse velocities are shown with the black stripes. The probability density function of the observed transverse velocities is shown in red. The model assuming a power-law distribution (see Supplementary Information) is shown in blue. \textbf{b}. The distribution of the distance statistic of the simulated model sample is shown in grey. The dashed black line indicates the 95\% threshold of the sample. The dashed red line shows the distance statistic for the observed transverse velocities ($p$-value = 0.27).}
\label{fig:contest_vt}
\end{figure}

We further explored the inherent structure of the dataset by employing K-means clustering \cite{MacQueen1967}, a widely used technique for partitioning data into distinct groups based on similarity. The objective behind using K-means clustering was to investigate the presence of distinct groupings within the data that could correspond to the two expected classes of runaway stars. These classes are theorised based on the result that the kinematic age distribution shows two distinct ejection events, one of which shows runaways being ejected anisotropically. By setting the number of clusters to two in the K-means algorithm, we aimed to identify whether the dataset naturally clusters into these hypothesised classes. This method was applied to the key quantities that are critical to understanding the dynamical properties and evolutionary states of runaway stars. These quantities, which include the kinematic age, transverse velocity, ejection angle, and evolutionary age, serve as a multidimensional feature space for clustering.

The clustering process began with standardising the four quantities to ensure that each feature contributes equally to the distance calculations, essential for the effectiveness of K-means. The algorithm then iteratively assigns each data point to one of two clusters based on the shortest distance to the cluster centroids, which are initially chosen at random. Through successive iterations, these centroids are recalculated and data points reassigned until the positions of the centroids stabilise, indicating that the clusters are as distinct as possible given the input data.

The two clusters show differences in the evolutionary age, kinematic age, and ejection angle distribution (Sup. Fig.~\ref{fig:runaways_kmeans}). One of the two clusters contains 12 runaways, of which 11 have kinematic ages lower than 1.0 Myr, evolutionary ages larger than 2 Myr and ejection angles between 50-150 deg. The other cluster contains nine runaways and shows a more uniform distribution in kinematic age. It contains runaways with evolutionary ages between 0-6 Myr, of which eight are between 0-2.5 Myr. The ejection angles are also more uniformly distributed between 50-350 deg. This provides the statistical basis for our conclusion that two physically different waves of runaways originate from R136.

\renewcommand{\figurename}{Sup. Fig.}
\renewcommand\thefigure{9}
\begin{figure}[h]
\centering
\includegraphics[width=0.99\linewidth]{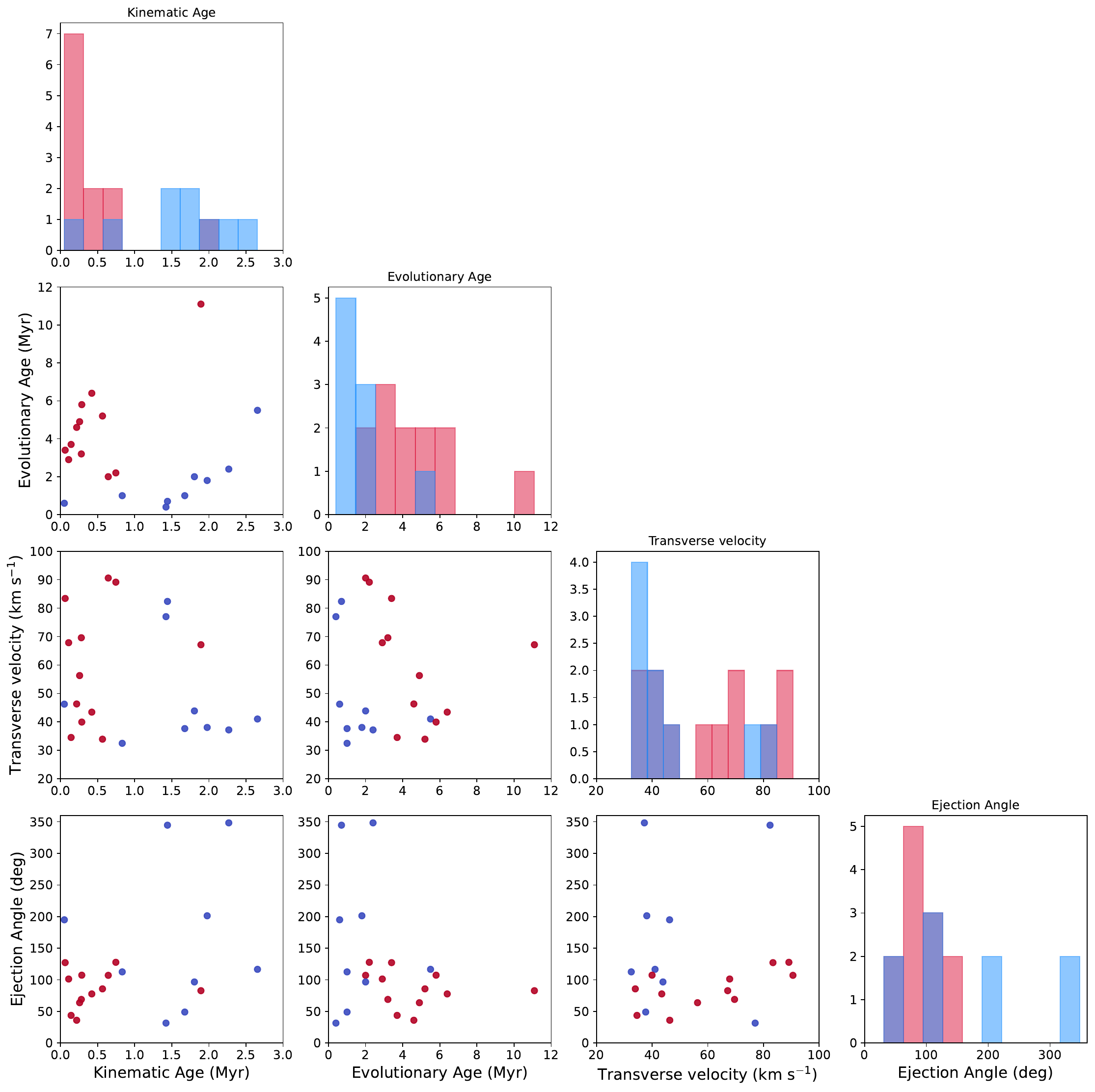}
\caption{\textbf{$\vert$ Corner plot of the K-means clustering assuming two clusters in the runaway properties}. The runaway parameters considered are the kinematic age, evolutionary age, transverse velocity, and ejection angle. The first cluster shown in red is distributed differently in the evolutionary age, ejection angle, and kinematic age space from the second cluster shown in blue. The diagonals show the distribution of the two different clusters. The off-diagonals show the scatter of the two clusters.}
\label{fig:runaways_kmeans}
\end{figure}

\renewcommand{\tablename}{Sup. Table}
\renewcommand\thetable{1}
\begin{sidewaystable}[h]
\caption{\textbf{$\vert$ Runaways coming from R136 in the past 3 Myr}. The uncertainty on the data is denoted as a 1$\sigma$ confidence interval.}
\label{tab:kinematic_runaways}
\tiny
\begin{tabular*}{\linewidth}{l l l l l l l l l l}
\toprule%
Identifier & source\_id & Spectral type & $v_{\rm{T}}$ & $t_{\rm{kin}}$ & $\phi$ & log($L$/L$_{\odot}$) & $M_{\rm{pri}}$ & $t_{\rm{evo}}$ & Ref. \\
- & - & - & km s$^{-1}$ & Myr & deg & - & M$_{\odot}$ & Myr & - \\
\midrule%
\noalign{\smallskip}Melnick 34 & 4657685534828257792 & WN5h + WN5h & 46.3 $\pm$ 6.4 & 0.052$^{+0.008}_{-0.007}$ & 194$^{+7.9}_{-7.8}$ & 6.43 & 139.0$^{+21.0}_{-18.0}$ & 0.6$^{+0.4}_{-0.4}$ & \cite{Tehrani2019} \\
\noalign{\smallskip}VFTS 590 & 4657685530496770944 & B0.7 Iab & 83.4 $\pm$ 6.2 & 0.063$^{+0.005}_{-0.004}$ & 127.2$^{+4.1}_{-4.1}$ & 5.87 & 46.8$^{+6.5}_{-6.1}$ & 3.4$^{+0.3}_{-0.4}$ & \cite{Evans2015a,Schneider2018} \\
\noalign{\smallskip}Cl* NGC 2070 SMB 182 & 4657685534828070144 & O7 & 67.9 $\pm$ 11.6 & 0.11$^{+0.02}_{-0.02}$ & 101.3$^{+11.2}_{-11.5}$ & 4.94 & 21.0 & 2.9 & \cite{Castro2018,Castro2021} \\
\noalign{\smallskip}VFTS 484 & 4657685534828237056 & O6-7 V((n)) & 34.5 $\pm$ 11.5 & 0.14$^{+0.07}_{-0.03}$ & 43.7$^{+20.4}_{-19.1}$ & 5.41 & 31.0$^{+4.1}_{-2.7}$ & 3.7$^{+0.3}_{-0.3}$ & \cite{Sana2013,Walborn2014,Schneider2018} \\
\noalign{\smallskip}VFTS 512 & 4657679655028521472 & O2 V-III((f*)) & 32.3 $\pm$ 5.7 & 0.16$^{+0.03}_{-0.02}$ & 268.6$^{+10.5}_{-10.5}$ & 6.04 & - & - & \cite{Walborn2014,SabinSanjulian2017} \\
\noalign{\smallskip}VFTS 476 & 4657685534828224896 & O6 V((f)) & 29.0 $\pm$ 9.4 & 0.22$^{+0.10}_{-0.05}$ & 33.4$^{+20.0}_{-19.0}$ & 4.9 & 25.0 & - & \cite{vanGelder2020} \\
\noalign{\smallskip}VFTS 419 & 4657686290721647872 & O9: V(n) & 46.3 $\pm$ 9.9 & 0.22$^{+0.06}_{-0.04}$ & 36.1$^{+12.0}_{-11.8}$ & 5.07 & 25.0$^{+2.7}_{-3.4}$ & 4.6$^{+0.6}_{-0.5}$ & \cite{Walborn2014,Schneider2018} \\
\noalign{\smallskip}Cl* NGC 2070 SMB 322 & 4657686286410992768 & O8 & 56.3 $\pm$ 15.9 & 0.26$^{+0.10}_{-0.06}$ & 63.8$^{+17.4}_{-16.1}$ & 4.51 & 17.0 & 4.9 & \cite{Castro2018,Castro2021} \\
\noalign{\smallskip}VFTS 550 & 4657685569187909120 & O5 V((fc))z & 69.6 $\pm$ 11.0 & 0.28$^{+0.05}_{-0.04}$ & 69.0$^{+10.0}_{-9.5}$ & 5.2 & 29.6$^{+3.5}_{-2.7}$ & 3.2$^{+0.4}_{-0.4}$ & \cite{Walborn2014,Schneider2018} \\
\noalign{\smallskip}Cl* NGC 2070 SMB 146 & 4657685569187801984 & O9 & 40.0 $\pm$ 10.9 & 0.29$^{+0.11}_{-0.06}$ & 107.3$^{+15.7}_{-17.2}$ & 4.84 & 20.0 & 5.8 & \cite{Castro2018,Castro2021} \\
\noalign{\smallskip}VFTS 529 & 4657685564905471488 & O9.5 (n) & 43.4 $\pm$ 13.1 & 0.42$^{+0.17}_{-0.09}$ & 77.7$^{+20.3}_{-19.0}$ & 4.71 & 17.8$^{+0.4}_{-0.4}$ & 6.4$^{+1.2}_{-1.3}$ & \cite{Walborn2014,Schneider2018} \\
\noalign{\smallskip}VFTS 531 & 4657685667984551552 & O9.5 III:nn & 45.3 $\pm$ 11.2 & 0.56$^{+0.18}_{-0.11}$ & 79.3$^{+16.6}_{-16.2}$ & - & 16.5$^{*}$ & - & \cite{Walborn2014,RamirezAgudelo2017} \\
\noalign{\smallskip}VFTS 517 & 4657685564905335680 & O9.5 V-III((n)) & 33.9 $\pm$ 9.1 & 0.57$^{+0.19}_{-0.11}$ & 85.7$^{+17.7}_{-17.0}$ & 5.09 & 21.0$^{+1.5}_{-1.1}$ & 5.2$^{+0.3}_{-0.3}$ & \cite{Walborn2014,Schneider2018} \\
\noalign{\smallskip}VFTS 830 & 4657682854768769024 & O5-6 V(n)((f)) & 127.2 $\pm$ 7.7 & 0.59$^{+0.04}_{-0.03}$ & 157.6$^{+3.6}_{-3.6}$ & 5.11 & 34.2$^{*}$ & - & \cite{Walborn2014,Shenar2021} \\
\noalign{\smallskip}RMC 144 & 4657686222022977920 & WN5/6h + WN6/7h & 90.6 $\pm$ 6.8 & 0.64$^{+0.05}_{-0.04}$ & 107.1$^{+3.9}_{-3.9}$ & 6.44 & 111.0$^{+12.0}_{-12.0}$ & 2.0$^{+0.3}_{-0.3}$ & \cite{Shenar2021} \\
\noalign{\smallskip}VFTS 758 & 4657685977170493568 & WN5h & 89.1 $\pm$ 6.2 & 0.75$^{+0.06}_{-0.05}$ & 127.7$^{+3.4}_{-3.3}$ & 6.36 & 85.6$^{+15.2}_{-16.6}$ & 2.2$^{+0.4}_{-0.2}$ & \cite{Evans2011,Bestenlehner2014,Schneider2018} \\
\noalign{\smallskip}VFTS 479 & 4657686462520312064 & O4-5 V((fc))z + B: & 39.8 $\pm$ 10.6 & 0.82$^{+0.29}_{-0.17}$ & 69.7$^{+16.1}_{-15.7}$ & 5.14 & 32.0$^{+6.0}_{-5.0}$ & - & \cite{Walborn2014,Shenar2021} \\
\noalign{\smallskip}VFTS 682 & 4657685637907503744 & WN5h & 32.5 $\pm$ 10.4 & 0.83$^{+0.35}_{-0.19}$ & 112.6$^{+18.3}_{-17.2}$ & 6.51 & 137.8$^{+27.5}_{-15.9}$ & 1.0$^{+0.2}_{-0.2}$ & \cite{Bestenlehner2011,Schneider2018} \\
\noalign{\smallskip}Gaia DR3 4657636743986481664 & 4657636743986481664 & - & 160.3 $\pm$ 11.9 & 1.18$^{+0.09}_{-0.08}$ & 172.2$^{+5.2}_{-5.1}$ & - & - & - & - \\
\noalign{\smallskip}VFTS 760 & 4657685981504903552 & A9-F0 II & 55.1 $\pm$ 4.6 & 1.21$^{+0.11}_{-0.09}$ & 114.9$^{+4.2}_{-4.2}$ & 4.11 & - & - & \cite{Evans2011,Schneider2018} \\
\noalign{\smallskip}2MASS J05404275-6909037 & 4657681549098830080 & - & 118.8 $\pm$ 14.1 & 1.32$^{+0.17}_{-0.14}$ & 191.2$^{+6.5}_{-6.9}$ & - & - & - & - \\
\noalign{\smallskip}VFTS 779 & 4657685912785446528 & B1 II-III & 44.0 $\pm$ 8.3 & 1.40$^{+0.31}_{-0.21}$ & 134.9$^{+10.8}_{-10.1}$ & 4.73 & 12.0$^{+3.0}_{-2.0}$ & - & \cite{Shenar2022} \\
\noalign{\smallskip}VFTS 72 & 4657698454092124416 & O2 V-III(n)((f*)) & 77.0 $\pm$ 5.0 & 1.42$^{+0.10}_{-0.09}$ & 31.6$^{+3.7}_{-3.7}$ & 6.07 & 97.6$^{+22.2}_{-23.1}$ & 0.4$^{+0.8}_{-0.4}$ & \cite{Walborn2014,Schneider2018} \\
\noalign{\smallskip}VFTS 37 & 4657689657978343040 & B2 III: & 93.4 $\pm$ 13.9 & 1.42$^{+0.24}_{-0.18}$ & 316.0$^{+8.0}_{-8.3}$ & - & - & - & \cite{Dufton2013,Evans2015a} \\
\noalign{\smallskip}VFTS 16 & 4657690620070706432 & O2 III-If* & 82.4 $\pm$ 4.2 & 1.44$^{+0.07}_{-0.07}$ & 344.6$^{+2.6}_{-2.6}$ & 6.12 & 91.6$^{+11.5}_{-10.5}$ & 0.7$^{+0.1}_{-0.1}$ & \cite{Evans2010,Schneider2018} \\
\noalign{\smallskip}2MASS J05400849-6904035 & 4657684224817613312 & - & 71.5 $\pm$ 13.5 & 1.61$^{+0.37}_{-0.25}$ & 145.1$^{+11.8}_{-11.3}$ & - & - & - & - \\
\noalign{\smallskip}Gaia DR3 4657635335237404288 & 4657635335237404288 & - & 172.9 $\pm$ 8.2 & 1.66$^{+0.08}_{-0.07}$ & 181.8$^{+2.6}_{-2.6}$ & - & - & - & - \\
\noalign{\smallskip}VFTS 252 & 4657686703059060224 & O8.5 Vz & 37.7 $\pm$ 8.2 & 1.68$^{+0.45}_{-0.29}$ & 49.0$^{+11.7}_{-12.0}$ & 4.73 & 21.8$^{+0.9}_{-0.9}$ & 1.0$^{+0.9}_{-0.8}$ & \cite{Walborn2014,Schneider2018} \\
\botrule%
\end{tabular*}
\end{sidewaystable}

\renewcommand{\tablename}{Sup. Table}
\renewcommand\thetable{1}
\begin{sidewaystable}[h]
\caption{Continued.}
\tiny
\begin{tabular*}{\linewidth}{l l l l l l l l l l}
\toprule%
Identifier & source\_id & Spectral type & $v_{\rm{T}}$ & $t_{\rm{kin}}$ & $\phi$ & log($L$/L$_{\odot}$) & $M_{\rm{pri}}$ & $t_{\rm{evo}}$ & Ref. \\
- & - & - & km s$^{-1}$ & Myr & deg & - & M$_{\odot}$ & Myr & - \\
\midrule%
\noalign{\smallskip}2MASS J05410904-6857342 & 4657731336333869184 & - & 128.9 $\pm$ 12.8 & 1.68$^{+0.18}_{-0.15}$ & 151.6$^{+6.5}_{-6.3}$ & - & - & - & - \\
\noalign{\smallskip}Gaia DR3 4657670347821452416 & 4657670347821452416 & - & 121.0 $\pm$ 17.2 & 1.70$^{+0.28}_{-0.21}$ & 343.0$^{+7.2}_{-7.1}$ & - & - & - & - \\
\noalign{\smallskip}SK -68 137 & 4657783322622150784 & O2-3 III(f*) & 111.0 $\pm$ 4.9 & 1.72$^{+0.07}_{-0.07}$ & 75.8$^{+2.1}_{-2.2}$ & - & - & - & \cite{Walborn1995} \\
\noalign{\smallskip}2MASS J05364212-6904026 & 4657692956532850176 & - & 87.9 $\pm$ 6.5 & 1.75$^{+0.14}_{-0.12}$ & 12.4$^{+4.6}_{-4.5}$ & - & - & - & - \\
\noalign{\smallskip}Gaia DR3 4657700206438680832 & 4657700206438680832 & - & 109.7 $\pm$ 11.2 & 1.79$^{+0.20}_{-0.16}$ & 44.7$^{+6.7}_{-7.0}$ & - & - & - & - \\
\noalign{\smallskip}VFTS 617 & 4657687733851287040 & WN5ha & 43.9 $\pm$ 8.6 & 1.80$^{+0.43}_{-0.29}$ & 96.6$^{+7.7}_{-7.6}$ & 6.29 & 88.4$^{+16.9}_{-15.8}$ & 2.0$^{+0.3}_{-0.3}$ & \cite{Foellmi2003,Schneider2018} \\
\noalign{\smallskip}OGLE LMC-ECL-20304 & 4657696220708164096 & - & 108.4 $\pm$ 13.9 & 1.84$^{+0.26}_{-0.20}$ & 8.8$^{+7.5}_{-7.2}$ & - & - & - & - \\
\noalign{\smallskip}2MASS J05374102-6905442 & 4657691960100539136 & - & 41.4 $\pm$ 9.6 & 1.84$^{+0.55}_{-0.34}$ & 6.5$^{+14.2}_{-14.1}$ & - & - & - & - \\
\noalign{\smallskip}VFTS 213 & 4657678559800125952 & B2 III:e & 43.0 $\pm$ 9.3 & 1.85$^{+0.52}_{-0.33}$ & 327.8$^{+12.0}_{-11.9}$ & - & - & - & \cite{Evans2015a,Villasenor2021} \\
\noalign{\smallskip}VFTS 295 & 4657688799003073408 & B0-0.5 V & 67.1 $\pm$ 12.2 & 1.89$^{+0.43}_{-0.29}$ & 82.8$^{+10.0}_{-10.1}$ & 4.09 & 11.2$^{+0.6}_{-0.7}$ & 11.1$^{+1.8}_{-1.8}$ & \cite{Evans2015a,Schneider2018} \\
\noalign{\smallskip}2MASS J05385591-6851492 & 4657782841585862528 & - & 103.0 $\pm$ 6.1 & 1.96$^{+0.12}_{-0.11}$ & 90.6$^{+2.8}_{-2.8}$ & - & - & - & - \\
\noalign{\smallskip}VFTS 797 & 4657676360777000320 & O3.5 V((n))((fc)) & 38.1 $\pm$ 6.5 & 1.98$^{+0.40}_{-0.28}$ & 201.3$^{+9.1}_{-9.6}$ & 5.6 & 48.6$^{+7.2}_{-5.4}$ & 1.8$^{+0.3}_{-0.3}$ & \cite{Walborn2014,Schneider2018} \\
\noalign{\smallskip}VFTS 772 & 4657688146168391936 & B3-5 V-III & 64.3 $\pm$ 17.3 & 2.04$^{+0.73}_{-0.42}$ & 88.8$^{+11.8}_{-11.9}$ & 3.41 & 5.1$^{\dagger}$ & - & \cite{Dufton2013,Evans2015a,Dufton2018} \\
\noalign{\smallskip}MCPS 084.44781-69.30846 & 4657653786424620800 & O2-3 V-III((f*)) & 92.3 $\pm$ 4.1 & 2.04$^{+0.09}_{-0.09}$ & 281.7$^{+3.4}_{-3.4}$ & - & - & - & \cite{Evans2015b} \\
\noalign{\smallskip}2MASS J05392689-6914056 & 4657675634882644864 & - & 58.9 $\pm$ 15.2 & 2.05$^{+0.67}_{-0.40}$ & 240.5$^{+13.1}_{-12.6}$ & - & - & - & - \\
\noalign{\smallskip}VFTS 65 & 4657691891381055232 & O8 V(n) & 37.2 $\pm$ 10.5 & 2.27$^{+0.81}_{-0.48}$ & 348.3$^{+15.6}_{-15.7}$ & 4.8 & 22.4$^{+1.8}_{-1.6}$ & 2.4$^{+0.9}_{-1.4}$ & \cite{Walborn2014,Schneider2018} \\
\noalign{\smallskip}BI 264 & 4657778649697948032 & O5/6 III(f) & 60.7 $\pm$ 5.9 & 2.41$^{+0.25}_{-0.20}$ & 134.1$^{+5.4}_{-5.2}$ & - & 37.4$^{*}$ & - & \cite{Evans2015b} \\
\noalign{\smallskip}Gaia DR3 4657283216685753216 & 4657283216685753216 & - & 187.0 $\pm$ 14.7 & 2.43$^{+0.21}_{-0.18}$ & 292.6$^{+3.9}_{-3.9}$ & - & - & - & - \\
\noalign{\smallskip}VFTS 148 & 4657698557171351424 & O9.7 II-III(n) & 42.6 $\pm$ 11.2 & 2.49$^{+0.80}_{-0.50}$ & 41.9$^{+14.8}_{-15.8}$ & - & 20.6$^{*}$ & - & \cite{Walborn2014} \\
\noalign{\smallskip}Gaia DR3 4657633750417399808 & 4657633750417399808 & - & 89.3 $\pm$ 13.3 & 2.51$^{+0.42}_{-0.32}$ & 198.2$^{+7.9}_{-8.4}$ & - & - & - & - \\
\noalign{\smallskip}KWV2015 J054212.76-685803.6 & 4657732195327468800 & A1 Ib / B8 III: & 114.4 $\pm$ 5.7 & 2.51$^{+0.13}_{-0.11}$ & 151.8$^{+3.3}_{-3.2}$ & - & - & - & \cite{Kamath2015} \\
\noalign{\smallskip}VFTS 33 & 4657690551351250560 & B1-1.5 V & 40.5 $\pm$ 11.9 & 2.64$^{+1.07}_{-0.61}$ & 316.2$^{+14.9}_{-15.8}$ & 4.26 & 11.8$^{\dagger}$ & - & \cite{Evans2015a,Villasenor2021} \\
\noalign{\smallskip}VFTS 843 & 4657684534063166848 & O9.5 IIIn & 41.0 $\pm$ 11.7 & 2.66$^{+0.98}_{-0.58}$ & 116.6$^{+14.0}_{-12.7}$ & 4.44 & 15.4$^{+0.8}_{-0.8}$ & 5.5$^{+1.0}_{-1.3}$ & \cite{Walborn2014,Schneider2018} \\
\noalign{\smallskip}2MASS J05383911-6934004 & 4657643615940780544 & - & 148.1 $\pm$ 13.2 & 2.66$^{+0.25}_{-0.22}$ & 262.6$^{+4.6}_{-4.6}$ & - & - & - & - \\
\noalign{\smallskip}2MASS J05403024-6909512 & 4657681613534527616 & - & 52.6 $\pm$ 11.5 & 2.67$^{+0.71}_{-0.46}$ & 185.5$^{+11.1}_{-11.9}$ & - & - & - & - \\
\noalign{\smallskip}2MASS J05401782-6911526 & 4657681132440914048 & - & 51.6 $\pm$ 10.3 & 2.76$^{+0.69}_{-0.45}$ & 189.2$^{+10.8}_{-11.5}$ & - & - & - & - \\
\noalign{\smallskip}Gaia DR3 4657680861904135296 & 4657680861904135296 & - & 62.5 $\pm$ 13.6 & 2.81$^{+0.76}_{-0.49}$ & 193.0$^{+13.0}_{-13.8}$ & - & - & - & - \\
\botrule%
\end{tabular*}
\footnotetext{$^{*}$Primary mass estimate based on O spectral type \citep{Martins2005}.}
\footnotetext{$^{\dagger}$Primary mass estimate based on B dwarf spectral type \citep{Pecaut2013}.}
\end{sidewaystable}

\end{appendices}

\end{document}